# Extracting Damping Ratio and Shear Wave Velocity through Joint Inversion of Rayleigh Wave Attenuation and Phase Velocity Data

Mrinal Bhaumik, Ph.D.,[1] and Brady R. Cox, Ph.D., P.E.
Department of Civil and Environmental Engineering, Utah State University, Logan, Utah, USA

[1]Corresponding author : mrinal.bhaumik@usu.edu

**Abstract**

Methods for reliable in-situ estimation of small-strain damping ratio remain far less developed than those for in-situ shear wave velocity ($V_s$) measurements. For surface wave-based approaches, challenges with extracting damping ratios are caused by both difficulties associated with developing wave attenuation coefficients from recorded wavefields and complexities involved when interpreting their multimodal irregularities during inversion. This study presents a joint inversion framework for extracting shear damping ratio ($D_{s0}$) and $V_s$ profiles from Rayleigh-wave attenuation and phase velocity data using a viscoelastic predominant-mode approach. Unlike conventional inversion, which requires explicit identification and indexing of individual modes, the predominant-mode formulation identifies the mode contributing most strongly to the measured wavefield at each frequency, allowing multimodal attenuation and dispersion information to be incorporated without explicit modal indexing. This approach has significant benefits for inverting attenuation coefficients, whose modal behavior is much more irregular than phase velocity. A coupled two-stage inversion strategy is adopted to jointly estimate the $V_s$ and $D_{s0}$ profiles while addressing their different sensitivities to phase velocity and attenuation. The predominant-mode attenuation concept is first validated against two synthetic ground models of increasing complexity. The synthetic inversion results demonstrate the framework's ability to recover the $D_{s0}$ and $V_s$ profiles and capture modal transitions associated with a low-velocity layer. The framework is then applied to active-source field data from Mirandola, Italy, where the fundamental mode dominates, and Bird's Landing, California, where a soft peat layer beneath a thin crust produces significant higher-modes. At both sites, the inverted $V_s$ profiles agreed well with independent site characterization data, while the damping estimates exhibit physically plausible trends and provide site-specific information that complements empirical estimates. The proposed framework offers a practical tool for non-invasive in-situ estimation of subsurface damping and stiffness.

***Keywords**:* Small-strain damping ratio; Rayleigh-wave attenuation; Joint inversion.

## 1. Introduction

An accurate assessment of small-strain shear modulus ($G_0$) and small-strain damping ratio ($D_0$) of subsurface materials is essential for dynamic analyses such as seismic site response, soil-structure interaction, and vibration assessments for geotechnical infrastructure (Kramer and Stewart 2024). The small-strain shear modulus $G_0$ is directly related to shear wave velocity ($V_s$) through $G_0 = \rho V_s^2$, where $\rho$ is mass density. Consequently, $V_s$ has become a standard parameter for site characterization and can be measured in situ using methods ranging from invasive crosshole and downhole testing to non-invasive surface-wave techniques. The latter are particularly attractive for routine site characterization because they are rapid, economical, applicable over a broad range of investigation depths, and capable of sampling relatively large soil volumes (Nazarian et al. 1983; Park et al. 1999). Although both Rayleigh and Love waves can be utilized for surface wave analysis, Rayleigh waves are most commonly employed in practice (Foti et al. 2018). By measuring Rayleigh wave phase velocity ($V_R$) as a function of frequency, one can infer how $V_s$ varies with depth via an inversion analysis. This approach has been extensively developed over the decades and is now routinely used for in-situ $V_s$ characterization.

In contrast, reliable in-situ measurement of $D_0$ remains challenging (Parolai et al. 2022) and $D_0$ is therefore more commonly evaluated from laboratory tests, such as resonant-column and torsional shear, or estimated using empirical correlations (Darendeli 2001; Menq 2003). Although laboratory tests provide valuable information regarding the strain-dependent damping ratio, the small-strain properties may be strongly influenced by sample disturbance, stress relief during sampling, specimen preparation procedures, boundary conditions, drainage conditions, and scale effects. Moreover, the field wave attenuation is commonly attributed to three mechanisms: intrinsic material damping, geometric spreading, and apparent attenuation due to wave scattering (Zywicki 1999). However, the small-scale lab measured $D_0$ can only resolve the intrinsic damping. As a result, laboratory-derived damping may not adequately represent the effective attenuation behavior of the in-situ wavefield and may lead to an underestimation of damping when field-scale propagation effects are significant (Dawadi et al. 2026). This disparity between laboratory-based damping and field-scale attenuation underscores the need for in-situ $D_0$ estimation. The development of surface-wave-based approaches for in-situ $D_0$ measurements have, however, due to their complexity, lagged at least three decades behind the well-established methods for in-situ $V_s$ characterization. This motivates the present work to focus on developing a joint estimation framework for both $V_s$ and $D_0$ from surface wave measurements.

In a viscoelastic medium, $D_0$ can be represented by separate shear and compressional damping ratios, $D_{s0}$ and $D_{p0}$, respectively. Theoretically, $D_{s0}$ can be higher than, comparable to, or lower than $D_{p0}$ (Abbas et al. 2025). Because Rayleigh wave phase velocity is predominantly sensitive to $V_s$, the shear wave damping, $D_{s0}$ is the primary damping parameter of interest in the present study. Accordingly, while $D_0$ was used initially in the general discussion of small-strain damping, the subsequent formulation and inversion focus specifically on $D_{s0}$.

The in-situ estimation of $D_{s0}$ from surface waves relies on inverting measured attenuation properties of Rayleigh waves. Similar to how $V_s$ is evaluated from inverting phase velocity $V_R$ dispersion data, by evaluating Rayleigh wave attenuation coefficients ($\alpha_R$) as a function of frequency, it is possible to estimate the depth-dependent $D_{s0}$ (Lai 1998; Rix et al. 2000). Several approaches have been developed to extract $\alpha_R$ from surface-wave measurements, including amplitude regression of surface-wave traces (Lai et al. 2002), the circle-fit method (Verachtert et al. 2018a), wavefield decomposition techniques (Bergamo et al. 2023), and frequency-domain beamforming-based attenuation (FDBFA) (Aimar et al. 2024a). Abbas et al. (2025) further extended FDBFA to extract attenuation from the ambient noise passive wavefield. Therefore, it is now possible to merge attenuation data from both active and passive measurements, similar to what is commonly done for dispersion data. However, attenuation data can exhibit considerably more complex behavior than phase velocity dispersion data. Theoretical attenuation curves may cross one another and exhibit irregular frequency-dependent variations, making the identification of individual modal branches difficult. As an illustrative example, Fig. 1 presents the theoretical phase velocity dispersion and attenuation curves for a simple one-layer-over-half-space model consisting of a 10-m thick layer with $V_S$ of 200 m/s overlying a half-space with $V_S$ of 400 m/s, with $D_{S0}$ of 2% for both layers. While the corresponding phase-velocity curves remain separated (Fig. 1a), the attenuation curves intersect over several frequency ranges (Fig. 1b).

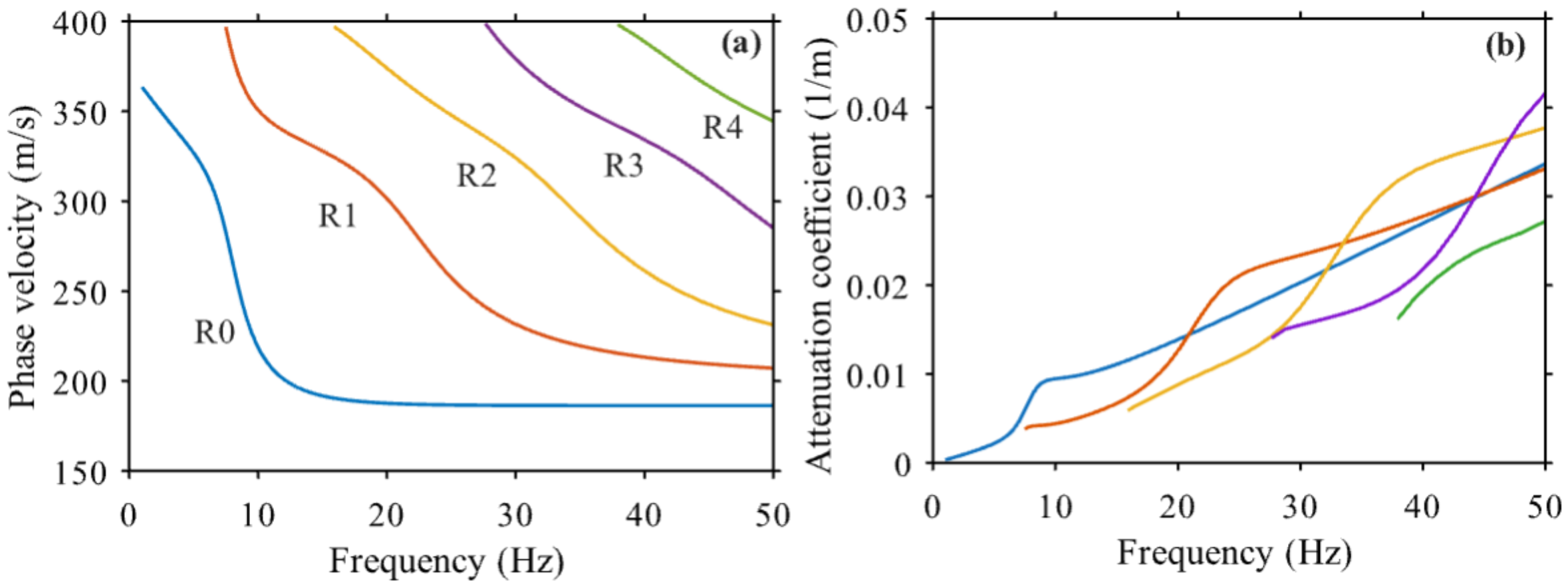

**Fig. 1.** Theoretical modal curves for a simple one-layer-over-half-space model: (a) Rayleigh-wave phase velocity dispersion curves and (b) attenuation coefficient curves. Note that R0 is the fundamental mode and R1, R2, R3, etc. are the sequential higher modes.

In addition, experimental attenuation estimates are generally more sensitive to factors such as near-field effects, geometric spreading, sensor coupling, noise contamination, source variability, and assumptions embedded in the processing workflow (Abbas et al. 2025; Parolai et al. 2022), resulting in greater uncertainty than phase velocity measurements. These characteristics become particularly important when attenuation data are incorporated into inversion, where the interpretation and indexing of individual attenuation modes can directly affect the estimated subsurface $D_{S0}$ .

Attenuation information can be incorporated into inversion either through an uncoupled approach, in which dispersion and attenuation data are inverted separately to estimate $V_S$ and $D_{S0}$, respectively (Rix et al. 2000), or through a fully coupled approach, where both parameters are simultaneously estimated within a single inversion framework (Lai et al. 2002). Conventionally, inversion is performed using the fundamental mode of Rayleigh waves, particularly at sites with monotonically increasing $V_S$ profiles, unless there is clear evidence of higher-mode excitation in the experimental dispersion data. However, higher-mode contributions can become important in the presence of velocity reversals, strong impedance contrasts, and other complex subsurface conditions. Aimar et al. (2024b) developed a Monte Carlo-based coupled inversion scheme capable of simultaneously inverting phase velocity and attenuation coefficient for multiple modes. However, for such multimodal inversions, the modes present in both the dispersion and attenuation data must be correctly identified and indexed.

While attenuation data modal transitions may be interpreted using the frequency bands associated with identified phase-velocity modal transitions, ambiguity in phase-velocity modal indexing can propagate into the interpretation of attenuation data and significantly affect the estimated $D_{S0}$. This issue becomes particularly important in the presence of velocity reversals, where higher modes may not be excited sequentially, making it difficult to establish a consistent correspondence between the identified dispersion modes and their associated attenuation modes. Similarly, strong impedance contrasts can cause modal energy to transition gradually from the fundamental mode to higher modes at relatively low frequencies through modal osculation or mode kissing (Bhaumik and Cox 2026a; Boaga et al. 2013; Kausel et al. 2015). Because the modal branches may become closely spaced during this transition, assigning the corresponding attenuation data to a specific phase velocity mode becomes difficult. These challenges motivate an inversion approach that can utilize multimodal attenuation information without requiring explicit identification and tracking of individual modes.

To address the ambiguity in phase velocity modal interpretation, Bhaumik and Cox (2026a) proposed a predominant-mode phase-velocity inversion strategy that does not require an analyst's judgment for mode indexing nor an unestablished fundamental-mode assumption. This predominant-mode inversion strategy is specifically advantageous when combining phase velocity dispersion data from multiple active shots with ambient noise dispersion data to generate a target phase velocity dataset for inversion, and where effective mode applications become challenging due to unknown and multiple source conditions. Bhaumik and Cox (2026a) demonstrated that their predominant-mode inversion successfully worked in several numerical and real-world examples involving complex modal osculation, and the method correctly identified the appropriate modal index based on the vertical surface modal amplitude, enabling it to capture velocity contrasts and layer interfaces with better accuracy than traditional inversion approaches.

This study proposes a new framework for joint estimation of $V_S$ and $D_{S0}$ using predominant-mode phase velocity and predominant-mode attenuation. Similar to predominant-mode phase velocity, the predominant-mode attenuation is defined by selecting the attenuation curve corresponding to the mode exhibiting the maximum vertical surface amplitude. Consequently, the proposed approach automatically identifies the physically dominant mode and eliminates the need for subjective modal identification and tracking, which becomes particularly challenging for attenuation data due to frequent modal intersections and irregular behavior discussed earlier. The proposed framework is particularly well suited for integrating dispersion and attenuation measurements obtained from multiple surface wave acquisition configurations/approaches (e.g., active and passive methods) into a unified inversion target, as it does not require explicit source–receiver information. Combining dispersion and attenuation data from multiple active-source shot locations improves inversion robustness by accounting for measurement uncertainty and reducing the influence of source-specific effects. Furthermore, combining active- and passive-source data is helpful when seeking to extend the usable frequency bandwidth and improve depth resolution.

To establish the proposed methodology, a viscoelastic forward modeling framework is first developed to simultaneously generate both predominant-mode phase velocity and predominant-mode attenuation curves. Subsequently, the existence of predominant-mode attenuation is established and validated using numerical simulations for subsurface models reported in the literature. A comprehensive parametric study is then performed to evaluate the influence of $D_{s0}$ on $V_R$, $\alpha_R$, and modal amplitudes at the free surface. Based on these findings, a joint inversion framework is developed to simultaneously invert dispersion and attenuation measurements for the estimation of both $V_s$ and $D_{s0}$ profiles. The proposed inversion methodology is first evaluated using the same synthetic subsurface models employed for validating predominant-mode attenuation and is then applied to two field datasets. The results demonstrate the capability of the proposed framework to provide a physically consistent and practical approach for simultaneous subsurface characterization of shear stiffness and damping properties.

## 2. Viscoelastic Forward Problem: Determination of Predominant-Mode Phase Velocity and Attenuation

The behavior of Rayleigh waves in a vertically heterogeneous layered half-space can be investigated using several forward modeling approaches, including the Thomson–Haskell propagator matrix method (Thomson 1950), the dynamic stiffness-matrix method (Kausel and Roësset 1981), and the thin-layer method (TLM) (Kausel and Peek 1982). Among these approaches, the primary advantage of the TLM for solving viscoelastic forward problems lies in its formulation as an eigenvalue problem. This formulation enables the simultaneous computation of modal complex wavenumbers (eigenvalues) and corresponding mode shapes (eigenvectors) without requiring an explicit root-search procedure. The attenuation coefficient is subsequently obtained from the imaginary part of the complex wavenumbers. In contrast, secular-function-based root search methods typically involve a two-step procedure. First, modal wavenumbers are determined for the undamped case. Subsequently, damping ratio is gradually introduced, and root searching is performed in the complex plane to identify the corresponding complex wavenumbers (Bhaumik and Naskar 2024a; Verachtert et al. 2018b). This process can become computationally expensive and challenging, particularly for multiple modes. Furthermore, the capability of the TLM to directly provide theoretical mode shapes for both the vertical and radial components of Rayleigh waves offers an additional advantage. These modal eigenfunctions are particularly useful for evaluating modal amplitudes at the free surface, which is essential for identifying the physically dominant (predominant) mode. The detailed derivation of the TLM can be found in the literature, and only the necessary steps are discussed here. Following the formulation introduced by Kausel and Roësset (1981), the frequency-wavenumber domain equation of motion for horizontally propagating waves can be assembled into the compact form:

$$[k^2\mathbf{A} + ik\mathbf{B} + (\mathbf{C} - \omega^2\mathbf{M})]\boldsymbol{\phi}_R = 0 \tag{1}$$

where $\boldsymbol{\phi}_R$ is the right eigenvector representing mode shape, $\omega$ is the angular frequency, $k$ is the wavenumber, and $i = \sqrt{-1}$. The global matrices $\mathbf{A}, \mathbf{B}, \mathbf{C}$ and $\mathbf{M}$ are assembled from the contributions of individual thin elements that discretize each soil layer thickness ($h$) along the depth axis. The resulting displacement field contains coupled radial and vertical components, reflecting the elliptical particle motion characteristic of Rayleigh waves. Specifically, the x-components represent the horizontal in-plane displacement, whereas the z-components represent the vertical displacement. Arranging all horizontal components first, followed by the vertical ones, the global matrices are given as:

$$\mathbf{A} = \begin{bmatrix} \mathbf{A}_{xx} & \mathbf{0} \\ \mathbf{0} & \mathbf{A}_{zz} \end{bmatrix};\ \mathbf{B} = \begin{bmatrix} \mathbf{0} & \mathbf{B}_{xz} \\ -\mathbf{B}_{zx} & \mathbf{0} \end{bmatrix};\ \mathbf{C} = \begin{bmatrix} \mathbf{C}_{xx} & \mathbf{0} \\ \mathbf{0} & \mathbf{C}_{zz} \end{bmatrix};\ \mathbf{M} = \begin{bmatrix} \mathbf{M}_{xx} & \mathbf{0} \\ \mathbf{0} & \mathbf{M}_{zz} \end{bmatrix} \tag{2}$$

At the element level, the corresponding sub-matrices are defined as (Kausel and Peek 1982):

$$\begin{aligned} \mathbf{A}_{xx}^{(e)} &= (\lambda + 2\mu)\int \mathbf{N}^T\mathbf{N}\,dz, \mathbf{A}_{zz}^{(e)} = \mu\int \mathbf{N}^T\mathbf{N}\,dz, \mathbf{B}_{xz}^{(e)} \\ &= \lambda\int \mathbf{N}^T\mathbf{N}'\,dz - \mu\int \mathbf{N}'^T\mathbf{N}\,dz, \mathbf{B}_{zx}^{(e)} = \mathbf{B}_{xz}^{(e)T}, \mathbf{C}_{xx}^{(e)} = \mu\int \mathbf{N}'^T\mathbf{N}'dz, \mathbf{C}_{zz}^{(e)} \\ &= (\lambda + 2\mu)\int \mathbf{N}'^T\mathbf{N}'dz, \mathbf{M}_{xx}^{(e)} = \mathbf{M}_{zz}^{(e)} = \rho\int \mathbf{N}^T\mathbf{N}\,dz \end{aligned} \tag{3}$$

where $\mathbf{N}(\mathrm{z})$ denotes the interpolation (shape) functions and $\mathbf{N}'(\mathrm{z})$ their derivatives, $\rho$ is density, $\lambda = \rho(V_p^2 - 2V_s^2)$ is the 1st Lame's parameter, $\mu = \rho V_s^2$ is the 2nd Lame's parameter, $V_p$ is compressional wave velocity of the layers, and $V_s$ is shear wave velocity. To model damping in weakly dissipative geomaterials at small strains, the medium is commonly assumed to behave as a linear viscoelastic solid with complex-valued elastic properties. This approach avoids the need to specify explicit time-domain rheological models. It requires that strains remain below the linear cyclic threshold shear strain (Vucetic 1994) and that the damping be low, here taken as $D_{s0} \leq 0.1$. Within this range, the approximate expression shown below in Equation (4) reproduces phase velocity exactly and attenuation within about 1% of the exact viscoelastic solution (Lai and Rix 2002; Meza-Fajardo and Lai 2007). Assuming frequency-independent hysteretic damping, the complex compressional and shear wave velocities can be expressed as (Lai and Rix 2002):

$$V_\mathrm{p}^* = \frac{V_p[1 + iD_{p0}]}{[1 + D_{p0}^2]};\ V_s^* = \frac{V_s[1 + iD_{s0}]}{[1 + D_{s0}^2]} \tag{4}$$

where $D_{s0}$ and $D_{p0}$ denote the damping ratios associated with shear and compressional waves, respectively. These complex elastic parameters are subsequently used in element-level matrix expressions in Equation (3) for **A**, **B**, and **C** within each element. Consequently, the assembled global matrices become complex-valued, leading to complex modal solutions. As a result, Rayleigh waves propagating in viscoelastic media are characterized by complex wavenumbers with distinct real and imaginary components representing wave propagation and attenuation characteristics. The quadratic eigenvalue problem in Equation (1) can be reformulated as a generalized eigenvalue problem of the same size (Kausel and Peek 1982):

$$\left(k^2 \begin{bmatrix} \mathbf{A}_{xx} & 0 \\ \mathbf{B}_{xz}^T & \mathbf{A}_{zz} \end{bmatrix} + \begin{bmatrix} \mathbf{C}_{xx} - \omega^2 \mathbf{M}_{xx} & \mathbf{B}_{xz} \\ 0 & \mathbf{C}_{zz} - \omega^2 \mathbf{M}_{zz} \end{bmatrix}\right) \begin{Bmatrix} \boldsymbol{\phi}_{\mathbf{x}} \\ ik\boldsymbol{\phi}_{\mathbf{z}} \end{Bmatrix} = 0 \tag{5}$$

where, $\boldsymbol{\phi}_{\mathbf{x}}, \boldsymbol{\phi}_{\mathbf{z}}$ are horizontal and vertical displacement components of the right eigenvector. For viscoelastic media, each Rayleigh wave mode yields a wavenumber $k = k_R + ik_I$, with real part $k_R$ governing the phase velocity and imaginary part $k_I$ controlling amplitude decay. The corresponding phase velocity and attenuation coefficient are given by,

$$\begin{aligned} v &= \frac{\omega}{\mathrm{Re}(k)} = \frac{\omega}{k_R} \\ \alpha &= -\mathrm{Im}(k) = -k_I. \end{aligned} \tag{6}$$

This attenuation reflects energy dissipation within the viscoelastic layers and determines the spatial decay rate of Rayleigh-wave amplitude as it propagates along the surface. Because the material properties are complex-valued, the resulting phase velocity and attenuation characteristics become frequency-dependent, leading to dispersive behavior even in viscoelastic homogeneous media.

After obtaining eigenvalues and eigenvectors by solving Equation (5), the corresponding eigenvectors are normalized according to the modal orthogonality condition (Barbosa and Kausel 2012). At the free surface ($z = 0$), the normalized vertical component of mode $m$ is written as $\widehat{\boldsymbol{\phi}}_{z,m}(0;\ k_m, \omega)$, which represents the theoretical surface amplitude associated with that mode at frequency $\omega$. By evaluating these amplitudes for all modes, the predominant mode is defined as the one with the largest vertical amplitude at the surface (Bhaumik and Naskar 2024b),

$$k_{predominant}(\omega) = \arg\max_{k_m} \left|\widehat{\boldsymbol{\phi}}_{z,m}(0;\ k_m, \omega)\right| \tag{7}$$

Similarly, the predominant attenuation can be obtained as:

$$\alpha_{predominant}(\omega) = \arg\max_{\alpha_m} \left|\widehat{\boldsymbol{\phi}}_{z,m}(0;\ k_m, \omega)\right| \tag{8}$$

This formulation provides a physically consistent criterion for identifying the dominant wave mode directly from the theoretical wavefield response. Note that the presented formulation assumes that the recorded wavefield is predominantly governed by the vertical component of Rayleigh waves. Similar formulations can be developed for the in-line radial component and can be further extended to incorporate source–receiver directivity effects. Such extensions may provide additional flexibility for applications involving different acquisition geometries and wavefield characteristics but are beyond the scope of the present work.

## 3. Validation of Predominant Mode Attenuation Curve

To validate the proposed predominant-mode attenuation framework, the attenuation curves obtained from the developed formulation in Equation (8) are compared against the attenuation data extracted from numerical wavefield simulations. Aimar et al. (2024a) conducted numerical simulations on two synthetic ground models (Table 1) and estimated attenuation from the simulated waveforms using the FDBFA method. In the present study, the same synthetic subsurface models are adopted, and forward modeling is performed to compute the corresponding theoretical phase velocity and attenuation coefficients using Equation (6). The theoretical dispersion and attenuation curves are subsequently plotted together with the normalized modal amplitudes at the free surface, $\widehat{\mathbf{\Phi}}_{z,m}(0;\ k_m, \omega)$, representing the actual measurement location under field conditions. The attenuation coefficients obtained from numerical wavefield modeling by Aimar et al. (2024a) are then plotted together with the theoretical predominant mode curves for comparison.

**Table 1.** Ground Model parameters adopted to validate predominant mode attenuation (after Aimar et al. 2024a).

| | Thickness (m) | Shear wave velocity $V_s$ (m/s) | Poisson's ratio | Density (kg/m3) | Shear wave damping ratio, $D_{s0}$ in % |
|---|---|---|---|---|---|
| Ground Model I | 5 | 200 | 0.33 | 1800 | 3.5 |
| | 10 | 300 | 0.33 | 1800 | 3 |
| | 10 | 400 | 0.33 | 1800 | 2.5 |
| | – | 500 | 0.33 | 1800 | 2 |
| Ground Model II | 5 | 250 | 0.33 | 1800 | 2.5 |
| | 3 | 150 | 0.33 | 1800 | 4 |
| | – | 350 | 0.33 | 1800 | 1 |

Figures 2a and 2b show the phase velocity dispersion and attenuation curves, respectively, for Ground Model I. As the model exhibits a monotonically increasing velocity profile, the fundamental mode carries most of the surface-wave energy and, therefore, dominates the entire frequency range in both dispersion and attenuation behavior. The simulated attenuation data in Fig. 2b reported by Aimar et al. (2024a) show strong agreement with the theoretical attenuation curve for the highest-amplitude fundamental mode. Ground Model II is an irregularly dispersive profile that contains a low-velocity layer, resulting in modal energy transitions from the fundamental mode to subsequent higher modes, as shown in Fig. 2c. Although it visually appears as a single continuous mode above 60 Hz, a smooth transition occurs near 85 Hz between Modes R2 and R3. The attenuation curves shown in Fig. 2d further demonstrate the highly irregular and intersecting nature of theoretical attenuation behavior. Nevertheless, the attenuation values extracted from the numerical simulations by Aimar et al. (2024a) consistently follow the attenuation branch associated with the highest normalized modal amplitude. Minor discrepancies below approximately 10 Hz for both models are primarily attributed to near-field effects. Overall, the comparison demonstrates strong agreement between the theoretical predominant-mode formulation and the attenuation characteristics derived from numerical wavefield simulations, thereby validating the proposed framework's ability to capture predominant-mode attenuation behavior in complex layered media.

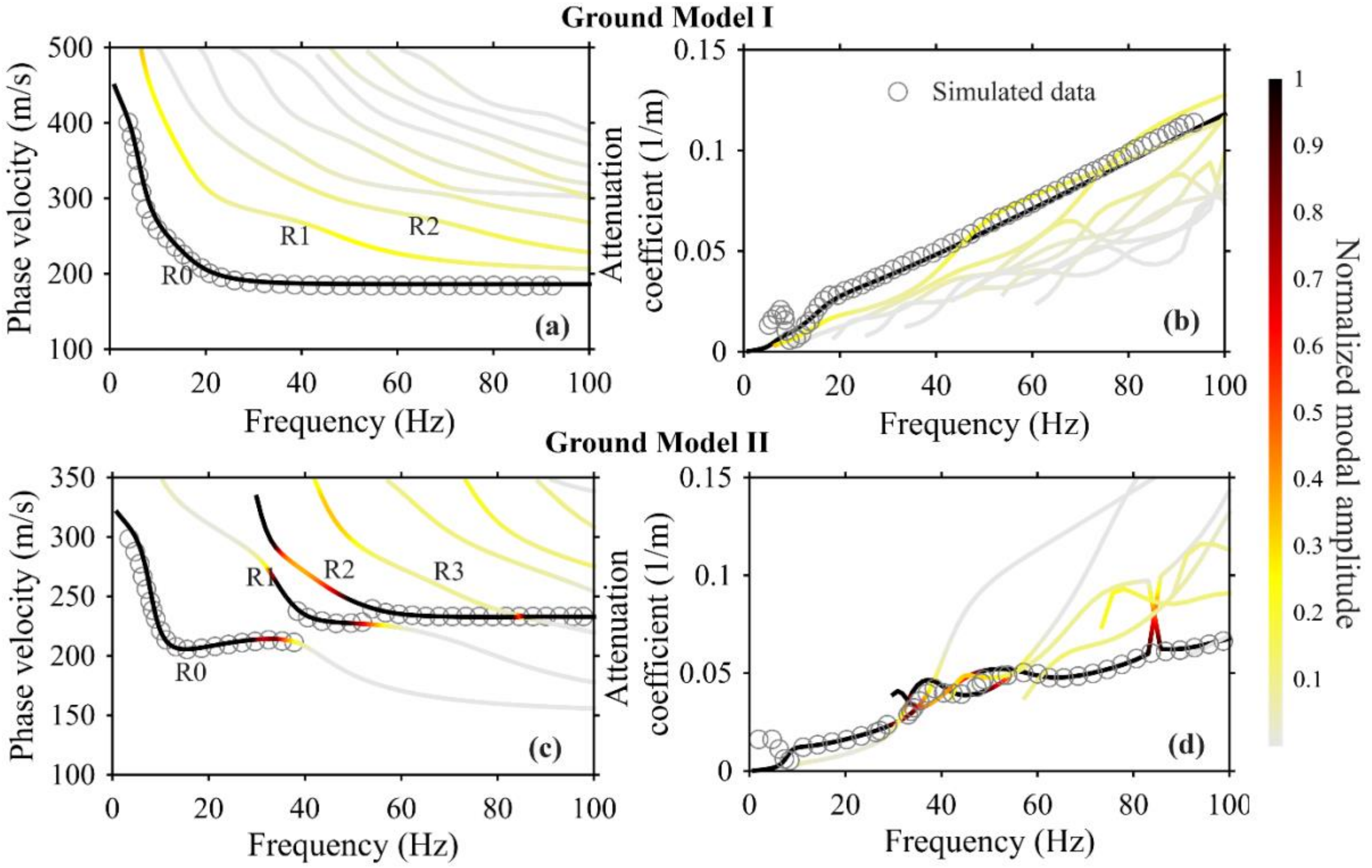

**Fig. 2.** Validation of the proposed predominant-mode attenuation framework through comparison with numerical wavefield simulation results reported by Aimar et al. (2024a): (a) theoretical phase velocity dispersion curves with normalized modal amplitudes for Ground Model I, (b) corresponding attenuation curves for Ground Model I, (c) theoretical phase velocity dispersion curves with normalized modal amplitudes for Ground Model II, and (d) corresponding attenuation curves for Ground Model II. The color of each modal curve indicates its normalized vertical surface modal amplitude (color bar). The simulated dispersion and attenuation data from Aimar et al. (2024a) are plotted using grey circle symbols.

# 4. Influence of Damping on Rayleigh Wave Phase Velocity, Attenuation, and Surface Modal Amplitude

For viscoelastic materials, the Rayleigh waves phase velocity and attenuation are a function of six primary layer parameters: $V_s, V_p, \rho, h, D_{s0}$ and $D_{p0}$. Theoretically, $D_{s0}$ can be higher than, comparable to, or lower than $D_{p0}$ (Abbas et al. 2025). Because Rayleigh-wave phase velocity is predominantly sensitive to $V_s$, the shear wave damping, $D_{s0}$ is the primary damping parameter of interest in the present study. For simplicity, the shear and compressional damping ratios are assumed to be equal ($D_{s0} = D_{p0}$) (Aimar et al. 2024b). This assumption is commonly adopted in surface-wave inversions since Rayleigh waves are primarily governed by shear deformation and exhibit much lower sensitivity to compressional damping compared with shear damping. For elastic media, the influence of the first four parameters on Rayleigh-wave dispersion characteristics has been extensively studied and is relatively well understood (Xia et al. 1999). In contrast, the influence of damping parameters on phase velocity, attenuation coefficient, and modal energy distribution in viscoelastic media remains comparatively less explored.

A parametric study is conducted to investigate the sensitivity of Rayleigh-wave phase velocity and attenuation coefficient to material damping. Understanding these sensitivities is important in identifying which parameters can be reliably constrained through inversion and provides a basis for developing an effective joint inversion strategy. In addition to phase velocity and attenuation coefficient, the influence of damping on normalized surface modal amplitudes is also examined. This is particularly important because variations in damping ratio may alter the distribution of modal energy and consequently affect predominant-mode behavior, modal transitions, and higher-mode excitation. Following the sensitivity analysis approach of Xia et al. (1999), the damping ratios of all layers are increased by 25%, and the resulting changes in phase velocity, attenuation coefficient, and normalized surface modal amplitudes of the first four modes are evaluated for the two synthetic ground models in Fig. 3.

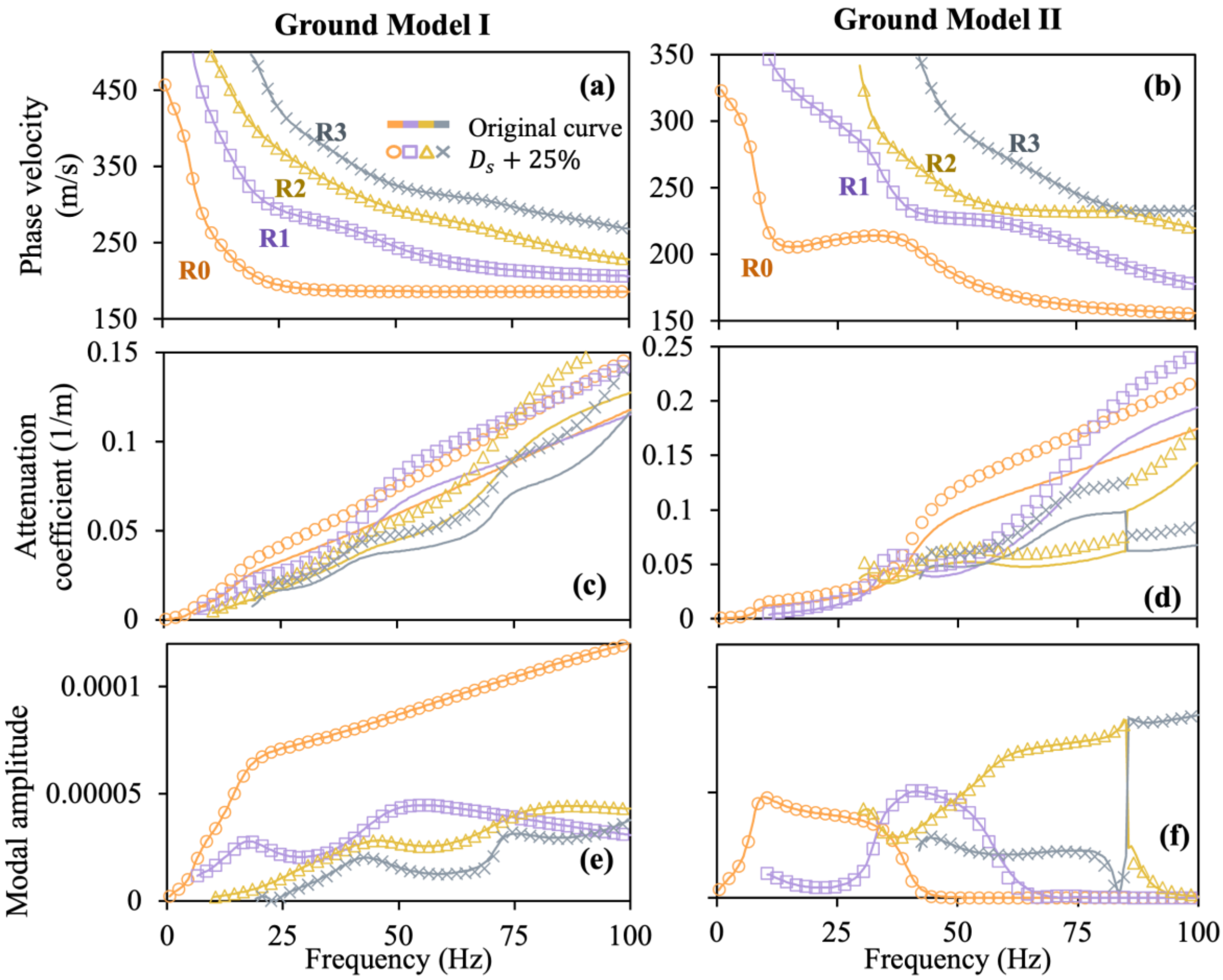

**Fig. 3.** Influence of damping on Rayleigh-wave phase velocity, attenuation, and surface modal amplitudes for Ground Models I and II: (a–b) phase velocity dispersion curves, (c–d) attenuation curves, and (e–f) surface modal amplitudes. Solid lines indicate the original model response, while markers represent the response after increasing the damping ratios of all layers by 25%.

Variations in damping ratio have a negligible influence on Rayleigh-wave phase velocity for both ground models. A 25% increase in damping ratio results in phase velocity changes of less than 0.2% across the investigated frequency range (Figs. 3a and 3b), indicating that phase velocity dispersion characteristics are primarily controlled by the elastic stiffness parameters rather than damping ratios. In contrast, attenuation coefficient exhibits significantly higher sensitivity to damping ratio. Increasing the damping ratios of all layers by 25% causes 24-30% variation in attenuation coefficients (Figs. 3c and 3d), which is also visually evident from the separation between the original and perturbed attenuation curves. The normalized modal amplitudes shown in Figs. 3e and 3f do not exhibit significant changes with increasing damping ratio. However, the overall modal energy distribution in the propagating wavefield may still be altered because different modes experience different attenuation during propagation. Therefore, damping ratio can influence the relative dominance of individual modes and modify modal energy transitions depending on the subsurface stratigraphy and propagation distance.

# 5. Joint Inversion Framework: Synthetic Examples

This section summarizes the key methodologies adopted for the two synthetic examples, including setting the inversion target, describing the parameter search bounds, selection of global inversion approach, definition of the objective function, and strategies for joint inversion.

## 5.1. Inversion target

The synthetic inversion targets were adopted from Aimar et al. (2024b), comprising dispersion and attenuation data and their associated uncertainty statistics, which were derived from numerically simulated wavefields for two ground models (Table 1). Aimar et al. (2024b) introduced Gaussian noise to represent measurement variability and modeled data variability using a lognormal distribution. For attenuation, the log-standard deviation was assigned as five times that used for phase velocity measurements. The details regarding the construction of target data and corresponding statistics can be found in Aimar et al. (2024b).

For Ground Model I, the target data used in the present study is identical to that adopted by Aimar et al. (2024b) and primarily corresponds to the fundamental mode (Figs. 4a and 4c), as expected for a regularly dispersive profile. For Ground Model II, although the low-velocity layer results in higher-mode responses at higher frequencies, only the fundamental-mode dispersion and attenuation data are available from Aimar et al. (2024b). However, Aimar et al. (2024a) provides additional dispersion and attenuation data associated with higher modes, as shown earlier in Figs. 2c and 2d. Since the objective of the present study is to evaluate the proposed predominant-mode approach under multimodal conditions, these higher-mode observations from Aimar et al. (2024a) were additionally incorporated into the Ground Model II target dataset (Figs. 4b and 4d). Therefore, the dispersion and attenuation data up to approximately 40 Hz, corresponding primarily to the fundamental mode, were adopted directly from Aimar et al. (2024b), while the higher-mode data from approximately 40-100 Hz were incorporated based on the modal information reported by Aimar et al. (2024a). For the additional higher-mode data, the uncertainty of the last available data point at approximately 40 Hz was applied across the extended higher frequency range, since corresponding uncertainty statistics were not available for these data. The inclusion of higher modes provides a more challenging scenario and allows assessment of the robustness and applicability of the proposed predominant-mode approach.

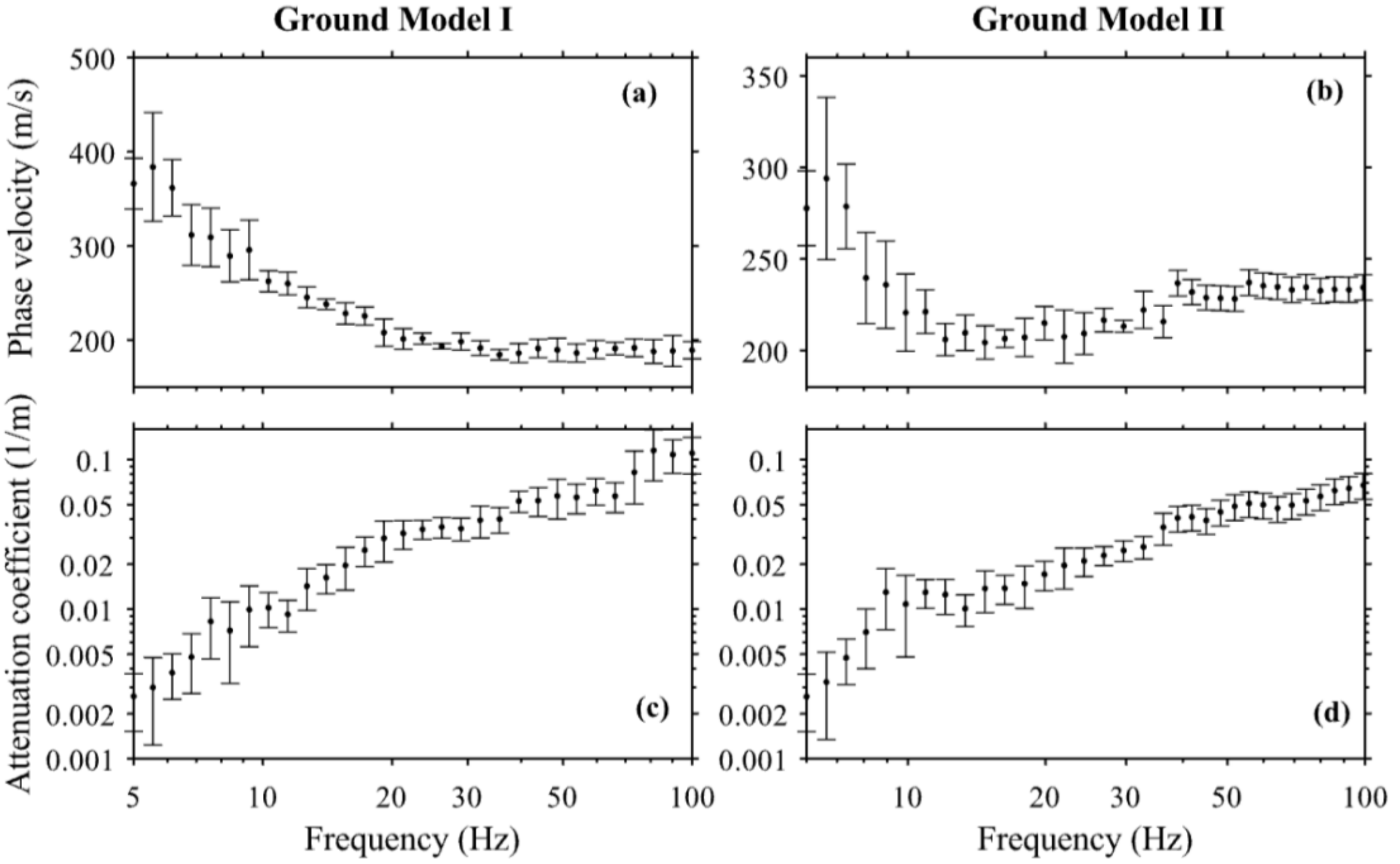

**Fig. 4.** Synthetic inversion targets used in the joint inversion. Panels (a) and (b) show phase velocity dispersion data, while panels (c) and (d) show attenuation coefficients for Ground Models I and II, respectively.

### 5.2. Inversion procedure: Differential Evolution (DE) algorithm

The predominant mode framework can be fit to any meta-heuristic global search algorithm to perform a joint inversion. This study used a differential evolution (DE)-based global search optimization (Storn and Price 1997), a population-based global optimization technique suitable for highly nonlinear, multi-dimensional problems by mimicking natural selection. It begins with a random population of candidate models, which are represented as a vector of free parameters and iteratively evolves through mutation, crossover, and selection operations. At each iteration, DE creates new trial vectors by taking the difference between two randomly selected individuals from the current population, and adding it to a third vector, a unique process known as differential mutation. These mutated components are then randomly mixed with the original vector's values during a crossover phase to form a new trial solution and it is accepted if they yield an improved objective function value. This iterative process allows the population to steadily converge toward the global optimal solution without requiring gradient information. To improve the convergence, an adaptive control parameter and $p$-best mutation strategy is used. In the beginning, an initial population of $N_p$ model vectors are randomly generated within the given parameter bounds, and corresponding objective functions are computed. At each generation, models are ranked based on their objective function value, and a subset corresponding to the top $p$ =10-20% of the population is identified as the $p$-best. A "current-to-$p$best/1" scheme (Zhang and Sanderson 2009) was used for mutation, where each parameter vector $\mathbf{x}_i^{(g)}$ at generation $g$ is perturbed according to:

$$\boldsymbol{v}_i^{(g)} = \mathbf{x}_i^{(g)} + F_i\left(\mathbf{x}_{pbest}^{(g)} - \mathbf{x}_i^{(g)}\right) + F_i\left(\mathbf{x}_{r1}^{(g)} - \mathbf{x}_{r2}^{(g)}\right), \tag{9}$$

where $\boldsymbol{v}_i^{(g)}$ is the mutated parameter vector, $\mathbf{x}_{pbest}^{(g)}$ is randomly selected from the $p$-best models, and $r1$ and $r2$ are indices. In the present implementation, the mutation factor $F_i$ was adaptively varied across generations to balance exploration and exploitation, decreasing linearly from $F_{\max} = 0.8$ to $F_{\min} = 0.3$.

The objective (i.e., misfit) function for phase velocity is defined as (Wathelet 2005):

$$m_d = \sqrt{\sum_{i=1}^{N_f} \frac{\left(\boldsymbol{V}_{\text{obs},i} - \boldsymbol{V}_{\text{predominant},i}\right)^2}{N_f \sigma_{v_i}^2}} \tag{10}$$

where $\boldsymbol{V}_{\text{obs},i}$ is the observed phase velocity at frequency $i$, $\boldsymbol{V}_{\text{predominant},i}$ is the modeled vertical-component predominant-mode velocity at the same frequency obtained from Equation (7), $N_f$ is the total number of frequency points, and $\sigma_{v_i}$ is the standard deviation associated with the observed phase velocity data at frequency $i$. Similarly, the objective function for attenuation can be defined as:

$$m_a = \sqrt{\sum_{i=1}^{N_f} \frac{\left(\boldsymbol{A}_{\text{obs},i} - \boldsymbol{A}_{\text{predominant},i}\right)^2}{N_f \sigma_{a_i}^2}} \tag{11}$$

where $\boldsymbol{A}_{\text{obs},i}$ is the observed attenuation coefficient at frequency $i$, $\boldsymbol{A}_{\text{predominant},i}$ is the modeled vertical-component predominant-mode attenuation at the same frequency obtained from Equation (8), and $\sigma_{a_i}$ is the standard deviation associated with the observed attenuation data at frequency $i$. For simplicity, the dispersion and attenuation data are sampled at the same frequency bins. The individual dispersion and attenuation misfits are combined to define the joint objective function:

$$m_{da} = m_d + m_a \tag{12}$$

The combined misfit provides a measure of the overall agreement between the observed and modeled dispersion and attenuation.

## 5.3. Joint inversion strategy

In the present formulation, the density of the layers is assumed to be constant, while $V_s, h, \nu,$ and $D_{s0}$ are treated as free parameters in the inversion. A straightforward inversion approach would be to perform a fully coupled joint optimization by minimizing the combined objective function, $m_{da}$. In such a formulation, all parameters evolve simultaneously under the influence of total misfit. However, sensitivity analysis presented in Section 4 showed that the phase velocity is only weakly sensitive to the damping parameter, $D_{s0}$, whereas the attenuation response is strongly sensitive to all the free parameters, including $V_s, h, \nu,$ and $D_{s0}$. As a consequence, directly minimizing the total misfit causes the attenuation misfit, $m_a$ to significantly influence the reconstruction of $V_s, h$ and $\nu$. This often slows the convergence of the joint inversion and increases the tendency of the optimization to become stagnated in local minima. To address this issue, an alternating two-stage optimization strategy is adopted in the present study. Instead of evolving all parameters simultaneously, the stiffness-parameters, $V_s$, $h$ and $\nu$ were evolved separately from damping ratio, $D_{s0}$ without fully decoupling them. During the first stage, only $m_d$ misfit was minimized to update $V_s, h$ and $\nu$ while keeping $D_{s0}$ unconstrained within the bound. Since $D_{s0}$ has a negligible influence on $m_d$, this stage effectively drives the stiffness-related parameters toward their optimal values based on $m_d$ information, while considering the viscoelastic effects. In the second stage, the combined objective function, $m_{da}$, is introduced and $D_{s0}$ is also included in the parameter updating process. During updating, the parameters are accepted only when $m_{da}$ improves. As the stiffness-related parameters have already been guided toward a near-optimal region during the first stage, the $m_a$ misfit primarily contributes to refining the damping ratio profile. Nevertheless, these parameters remain free and can continue to evolve if further improvements in the combined objective function are achieved. This strategy, therefore, represents a semi-coupled inversion framework, where phase velocity dispersion information first stabilizes the estimation of stiffness-related parameters before attenuation information is introduced to constrain damping. This approach reduces interference between the two objective functions, improves convergence stability, and provides a computationally practical solution for the simultaneous estimation of $V_s$ and $D_{s0}$. It should be noted that the proposed framework is not intended to represent a best solution strategy to the coupled inversion problem. Alternative formulations, such as fully-coupled multi-objective optimization or adaptive weighting, may offer improved convergence and computational efficiency. Nevertheless, the present strategy was found to provide a computationally practical and effective framework.

## 5.4. Inversion parameterization

A reasonable parameterization of the trial subsurface models is necessary to balance resolution and stability during inversion. Any available geological data or independent geotechnical survey often provide valuable information and should be considered in selecting the model layering (i.e., possible layer interface depths, number of layers, etc.). While the true number of layers is known for each synthetic ground model, we wanted to treat the inversion layering parameterizations as if we did not know the true number of layers, thereby reflecting more challenging and realistic conditions. In the absence of a priori information on the true number of subsurface layers, a systematic approach based on layering ratios (LR), as proposed by Cox and Teague (2016) and implemented in the *SWinvert* workflow (Vantassel and Cox 2021), can be used to investigate several layering parameterizations. More recently, Vantassel and Cox (2021) introduced an alternative parameterization known as layering by number (LN), which was shown to outperform LR in several applications. In the LN approach, the total number of layers are selected to discretize the subsurface model. The minimum thickness of each layer is constrained by the minimum resolvable wavelength, $\lambda_{\min}$, and is typically set to approximately $\lambda_{\min}/3$. The maximum investigation depth is controlled by maximum resolved wavelength, $\lambda_{max}$, and is generally taken to be between $\lambda_{max}/3$ to $\lambda_{max}/2$. As the true number of subsurface layers is generally unknown, relying on a single parameterization will underestimate epistemic uncertainty in the inversion results (Cox and Teague 2016). To account for this uncertainty, multiple LN parameterizations (4, 5, and 6) were considered in the present study, thereby avoiding dependence on a single layering scheme and allowing exploration of model uncertainty.

The $V_s$ bounds for each layer were defined based on the observed dispersion data. The lower bound was set at half of the minimum phase velocity, while the upper bound was selected as 1.5 times the maximum phase velocity. Velocity reversals are allowed only if there is visual evidence of low velocity layers in the target phase velocity dispersion data. Exciting higher modes and the presence of upward then downward trending portions of the dispersion data are some of the indications of velocity reversals in the subsurface. Mass density was assumed to be constant at 1800 kg/m$^3$, and Poisson's ratio was allowed to vary within the range of 0.25-0.35, representing un-saturated subsurface conditions.

For simplicity, $D_{p0}$ is assumed to be equal to $D_{s0}$. The search ranges of $D_{s0}$ are selected based on the expected material type and reported small-strain damping ratio values. Kokusho (2017) reported typical $D_{s0}$ values of approximately 3-5% for fine-grained soils and around 1% for coarse-grained soils. Darendeli (2001), based on laboratory test over a broad range of confining pressures and plasticity index, developed an empirical relationship showing that $D_{s0}$ for fine-grained soils generally falls around 1-2%, with values approaching 3% for highly plastic soils. Menq (2003) similarly provides empirical estimates for coarse-grained materials, which are generally less than 1%. However, attenuation obtained from field surface-wave measurements reflect not only intrinsic material damping but also scattering and other wave-propagation effects and can therefore yield larger damping ratio values. Therefore, a broader admissible $D_{s0}$ search ranges of 0.5–10% can be adopted to avoid overly restrictive bounds on the inversion. Although small-strain damping ratio exhibits substantial variability, empirical studies indicate that it depends systematically on soil type and state variables such as plasticity and confining stress (Darendeli 2001; Menq 2003). Several studies have shown that stiff, high-velocity materials generally exhibit lower small-strain material damping whereas softer soils are commonly characterized with higher small-strain material damping (Bergamo et al. 2023). Therefore, instead of fully allowing the $D_{s0}$ profile to vary completely independently between adjacent layers, a weak regularization is introduced to encourage consistency between the overall trend of $V_s$ and $D_{s0}$. A weighted Pool Adjacent Violators Algorithm (PAVA) (Leeuw et al. 2010) is employed to promote physically reasonable $D_{s0}$ distributions while allowing sufficient flexibility for the inversion to fit the data. This regularization does not enforce any relationship between $D_{s0}$ and $V_s$; instead, it discourages unrealistic random fluctuations in $D_{s0}$ between the adjacent layers.

This parameterization is consistent with the depth-dependent resolving capability of surface waves and provides a controlled, yet flexible, framework for inversion. Note that the adopted inversion parameterization represents one reasonable practical choice for the present study; however, alternative parameterizations may be adopted depending on site conditions and prior information.

# 6. Synthetic Inversion Results

## 6.1. Ground Model I

As shown in Table 1, Ground Model I is a three-layer monotonically increasing $V_s$ profile with interfaces located at 5, 15, and 25 m. The $D_{s0}$ parameterization limit for this model was set between 0.5-7 %. The joint inversion results for Ground Model I are presented in Fig. 5.

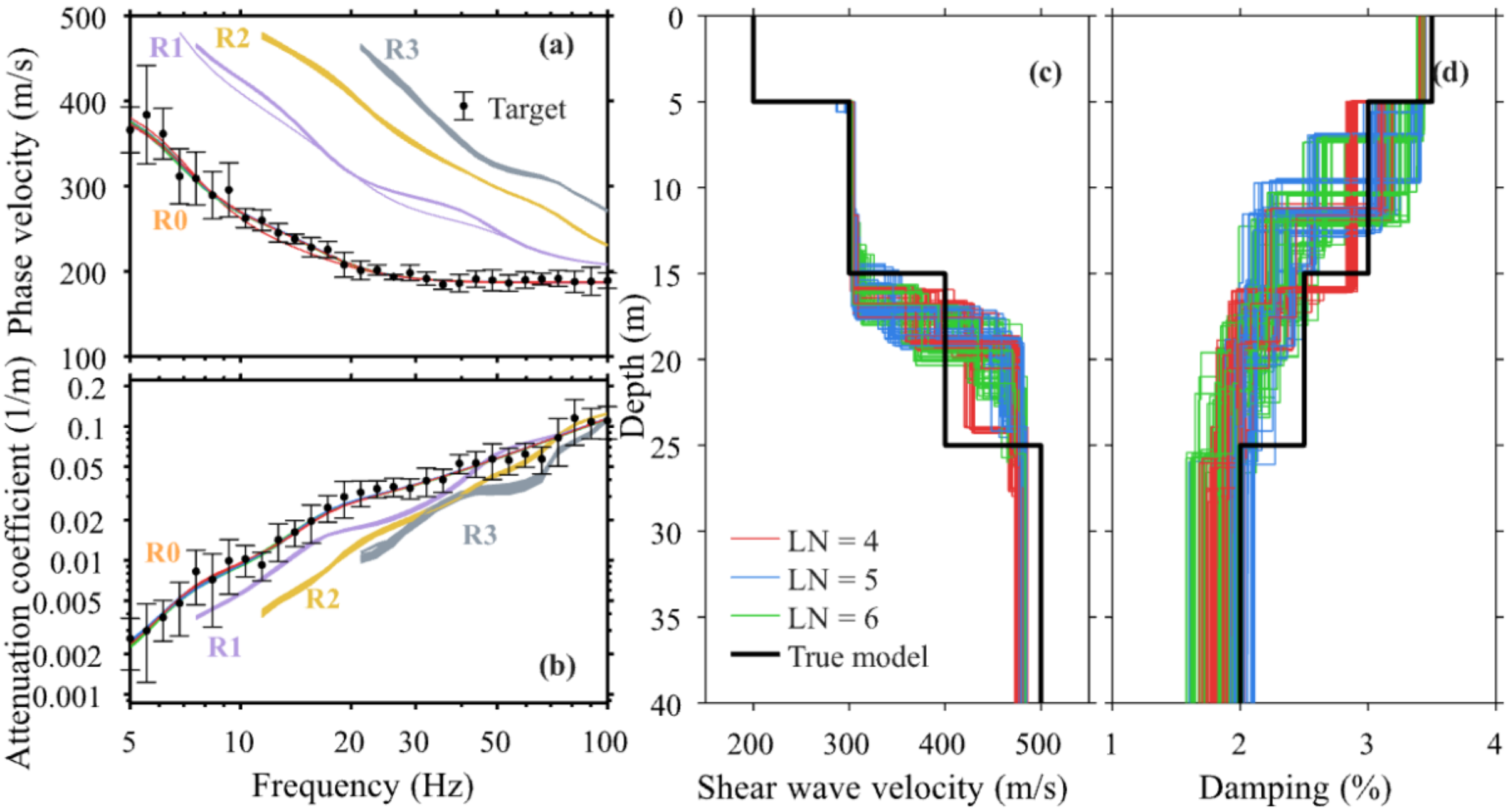

LN = 4 : [0.62 , 0.85]; LN = 5 : [0.83 , 0.85]; LN = 6 : [0.83 , 0.85]

**Fig. 5.** Results of the proposed joint inversion framework for Ground Model I. Panels (a) and (b) show the target phase velocity and attenuation data together with the theoretical modes (R0–R3) and the corresponding predominant-mode from the ten best-fitting solutions of each inversion run (five runs for each layer parameterization: LN = 4, 5, and 6), resulting in 150 solutions in total. Panels (c) and (d) present the corresponding reconstructed $\boldsymbol{V_s}$ and $\boldsymbol{D_{s0}}$ profiles, along with the true subsurface model. The ranges reported below the figure indicate the minimum and maximum total misfit values obtained for each layer parameterization.

For Ground Model I, the target phase velocity and attenuation data contain only the fundamental mode. Figures 5a and 5b show the target phase velocity and attenuation data together with the ten best-fitting solutions obtained from each of five inversion trials of three layering parameterizations. The theoretical modes (R0–R3) corresponding to the 150 inverted profiles are shown together with the predominant-mode response of each profile (shown as the red, blue, and green lines associated with LN = 4, 5, and 6, respectively). As observed, both the dispersion and attenuation targets are well-fit by the inversion results. The fitted predominant-mode responses consistently follow the fundamental mode (R0), which is expected for a regularly dispersive profile. The corresponding total misfit values for each LN parameterization are indicated in brackets at the bottom of the figure. The inverted $V_S$ profiles are shown in Fig. 5c together with the true model. All layering parameterizations successfully capture the overall velocity structure. The shear-wave velocities of the first and second layers as well as the half-space are recovered with good accuracy. Greater variability is observed in the estimated depth of the deepest interface at 25 m. Figure 5d presents the corresponding damping ratio profiles. Similar to the velocity inversion results, the inverted $D_{S0}$ profiles successfully reproduce the overall damping ratio structure of the model, although greater scatter is observed among the intermediate layers. This behavior is expected because attenuation measurements generally contain larger uncertainty than phase velocity measurements and because attenuation is influenced by a broader set of parameters. Nevertheless, the inversion consistently captures the overall damping ratio trends, demonstrating that the proposed framework can simultaneously recover both stiffness and damping ratio profiles with reasonable accuracy for this simple synthetic model.

### 6.2. Ground Model II

The joint inversion results for Ground Model II are presented in Fig. 6. This profile contains a low-velocity layer, resulting in irregular dispersion behavior and significant higher-mode contributions. Consequently, Ground Model II represents a more challenging inversion problem and provides a rigorous test of the proposed predominant-mode framework.

Figures 6a and 6b show the target dispersion and attenuation data together with the ten best-fitting predominant-mode solutions obtained from each of five independent inversion trials for each parameterization. Similar to Ground Model I, the theoretical modes (R0–R3) corresponding to the 150 inverted profiles are shown together with their predominant-mode responses. The inverted phase velocity and attenuation curves show good agreement with the target data across the entire frequency range. Notably, the inversion successfully reproduces the modal transitions associated with the low-velocity layer, demonstrating that the predominant-mode formulation can naturally consider mode jumps without requiring explicit modal identification. The corresponding inverted $V_S$ profiles are shown in Fig. 6c. All inversion runs successfully identify the presence of the low-velocity layer and recover the overall velocity structure with reasonable accuracy. The depths of the major interfaces are consistently estimated, while some variability is observed in the velocity contrast and thickness of the intermediate layers. Such variability is expected because multiple combinations of layer thickness and velocity can produce similar dispersion and attenuation responses, particularly in the presence of strong modal interference. Nevertheless, the recovered profiles consistently reproduce the key features of the true model. Figure 6d presents the corresponding damping ratio profiles. Similar to the velocity results, the inversion successfully captures the overall damping ratio structure and correctly identifies the higher damping associated with the low-velocity layer. The estimated damping ratio values remain physically consistent and follow the principal trends of the true model. Overall, the results demonstrate that the proposed joint inversion framework remains robust even for irregularly dispersive profiles exhibiting significant higher-mode contributions.

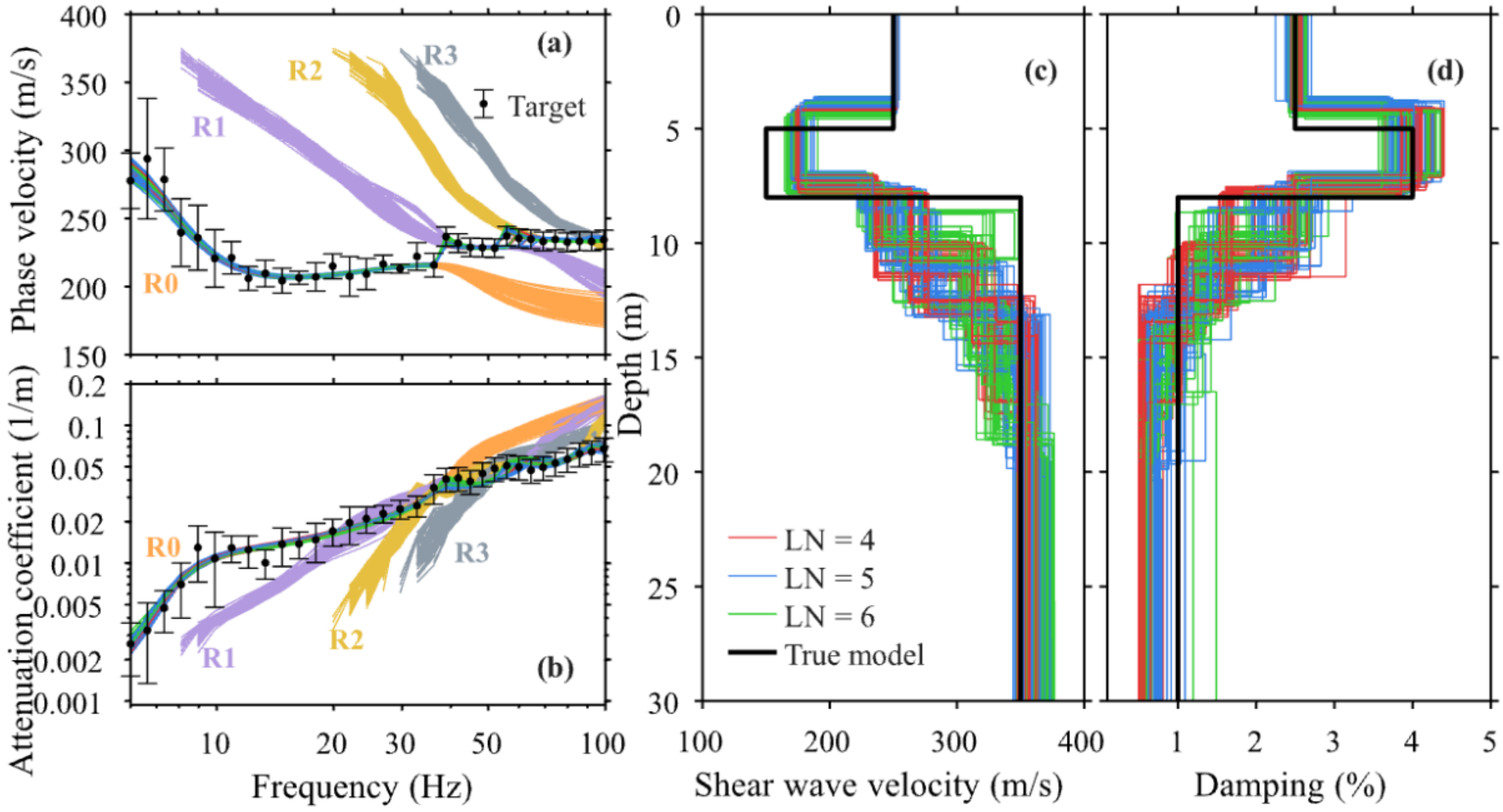

LN = 4 : [0.77 , 0.94]; LN = 5 : [0.79 , 0.96]; LN = 6 : [0.75 , 0.94]

**Fig. 6.** Results of the proposed joint inversion framework for Ground Model II. Panels (a) and (b) show the target phase velocity and attenuation data together with the theoretical modes (R0–R3) and the corresponding predominant-mode from the ten best-fitting solutions of each inversion run (five runs for each layer parameterization: LN = 4, 5, and 6), resulting in 150 solutions in total. Panels (c) and (d) present the corresponding reconstructed $\boldsymbol{V_s}$ and $\boldsymbol{D_{s0}}$ profiles, along with the true subsurface model. The ranges reported below the figure indicate the minimum and maximum total misfit values obtained for each parameterization.

## 7. Field Examples

Following the successful validation of the proposed joint inversion framework on synthetic datasets, the methodology is subsequently evaluated using two field case studies. The first dataset corresponds to Site A from Mirandola, Italy, acquired as part of the InterPACIFIC project (Garofalo et al. 2016b; a), while the second dataset corresponds to Site B located at Bird's Landing, near Sacramento, California (Panthi and Cox 2026) with the data publicly available through DesignSafe (Cox et al. 2026). For both sites, dispersion and attenuation characteristics were extracted from the measured wavefields using the FDBFA method (Aimar et al. 2024a).

Since direct field measurements of small-strain damping are rarely available in routine site investigations, current engineering practice commonly relies on empirical correlations to estimate the $D_{\min}$. Therefore, in addition to the inversion results shown below, damping ratio values estimated using the widely adopted relationship proposed by Darendeli (2001) are also evaluated and compared with the inverted damping ratio profiles. Darendeli's formulation accounts for the influence of mean effective confining pressure ($\sigma'_m$), overconsolidation ratio (OCR), plasticity index (PI), and loading frequency ($f$) in Hz:

$$D_{\min}(\%) = (0.8005 + 0.0129 \cdot \mathrm{PI} \cdot \mathrm{OCR}^{-0.1069}) \cdot {\sigma'_m}^{-0.2889} \cdot [1 + 0.2919 \cdot \ln(f)] \quad (13)$$

This relationship was developed using laboratory resonant column and torsional shear test data and the work by Darendeli (2001) has been widely adopted for the generation of modulus reduction and damping curves in equivalent-linear and nonlinear ground-response analyses. Comparison between the inverted damping ratio profiles and the empirical estimates provides a useful benchmark for evaluating the plausibility of the field-derived damping ratio values and assessing the potential of attenuation-based surface-wave inversion as a tool for in-situ damping characterization.

### 7.1. Site A: Mirandola

The Mirandola site has become a benchmark for the evaluation of surface-wave methods due to the availability of a comprehensive geotechnical and geophysical dataset (Garofalo et al. 2016b). The subsurface is characterized by alternating layers of alluvial silts, clays, and sandy deposits associated with the Po River basin (Garofalo et al. 2016a). The active-source surface-wave data used in this study were acquired using an array of 48 vertical-component geophones spaced at 2 m intervals. An example sledgehammer shot gather from a 14 m offset is shown with the corresponding dispersion and attenuation images extracted from the waveforms in Fig. 7.

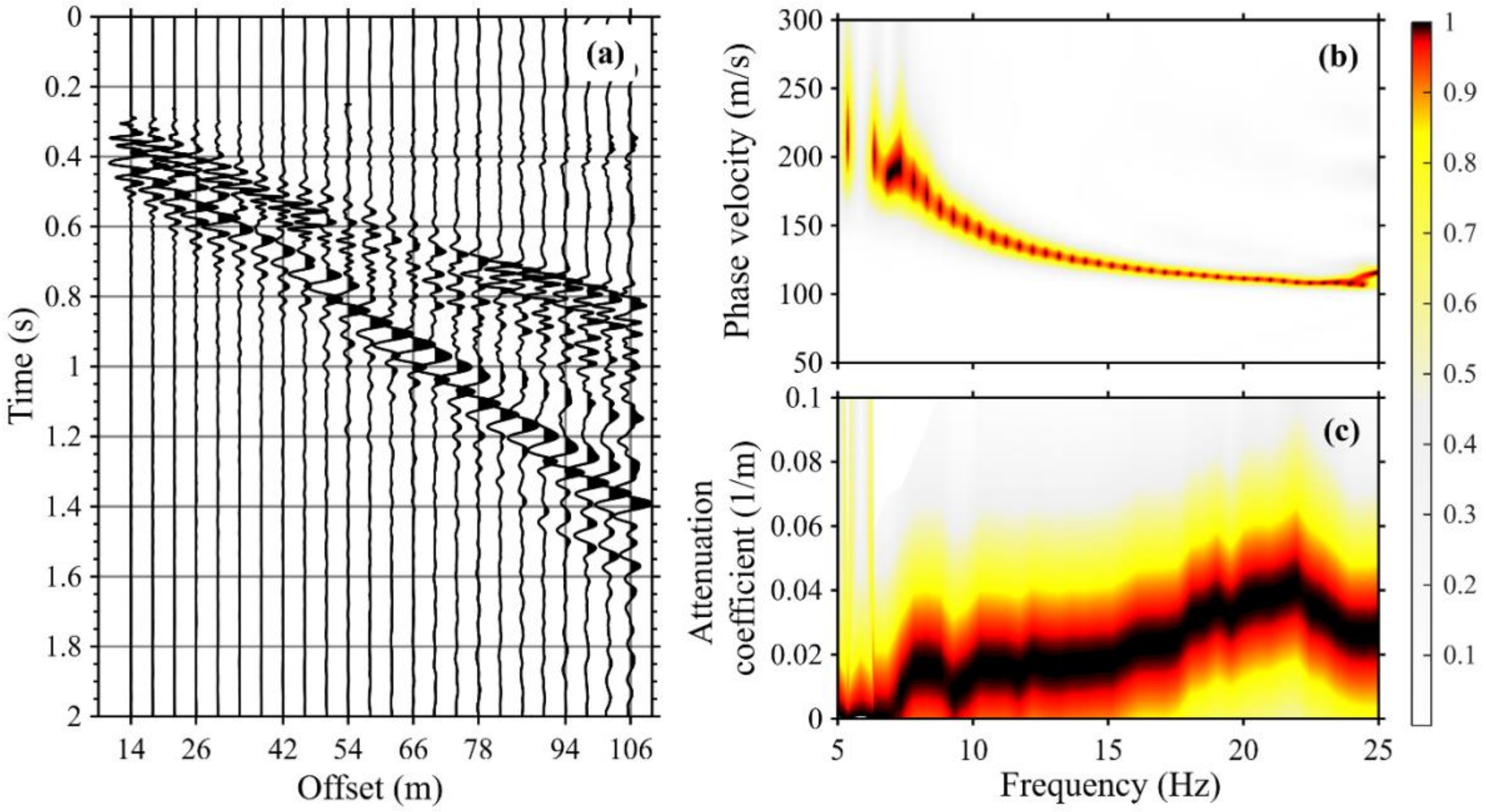

**Fig. 7.** Example 48-channel active shot waveforms for dispersion and attenuation processing at Site A: (a) recorded waveforms from a representative hammer shot at a 14-m offset, (b) corresponding Rayleigh-wave dispersion image, and (c) attenuation image.

The phase velocity dispersion image (Fig. 7b) reveals a dominant fundamental mode extending from approximately 5 to 25 Hz. The attenuation image obtained using the FDBFA method is shown in Fig.

7c. The attenuation measurements exhibit greater uncertainty and broader energy distributions than the corresponding phase velocity measurements. Nevertheless, the attenuation image displays a well-defined attenuation trend that can be used to constrain damping properties during inversion. The inversion targets were constructed by combining the dispersion and attenuation measurements extracted from all available active-source shot records, which consisted of shot offsets of 3.5 m and 4 m on one side of the array and 3 m and 14 m on the other side. The resulting phase velocity and attenuation target curves are shown in Figs. 8a and 8b, respectively. Similar to the synthetic examples, inversions were performed using LN values of 4, 5, and 6 to account for uncertainty associated with layer parameterization. The corresponding inversion results are presented in Fig. 8.

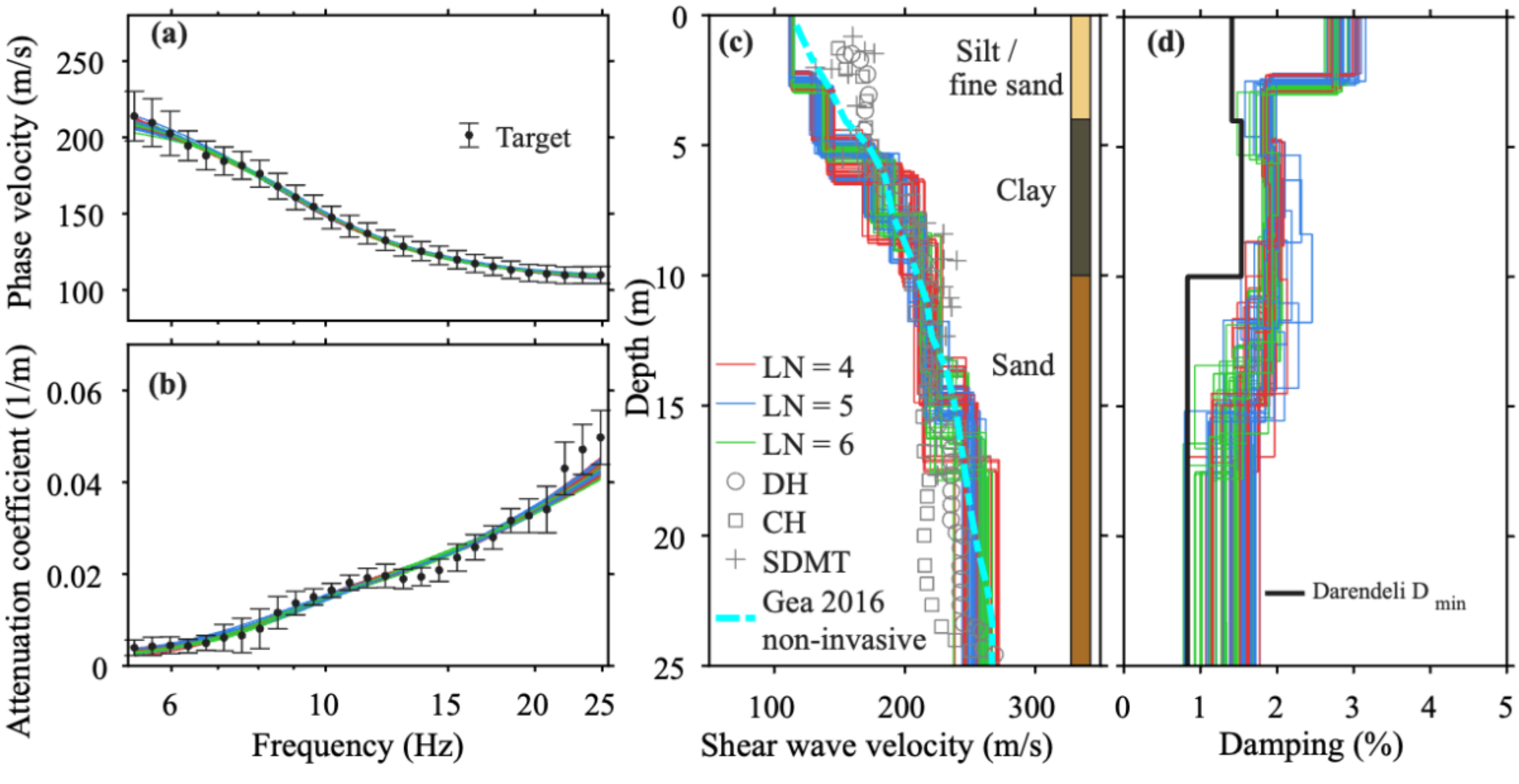

**Fig. 8.** Results of the proposed joint inversion framework for Site A. Panels (a) and (b) show the target dispersion and attenuation data together with the predominant-mode responses of the five best-fitting solutions from each inversion run (five runs for each layer parameterization: LN = 4, 5, and 6), resulting in 75 solutions in total. Panel (c) presents the inverted $\boldsymbol{V_s}$ profiles and those from downhole (DH), crosshole (CH), and seismic dilatometer testing (SDMT), as well as the overall non-invasive median Vs profile for the site reported by Garofalo et al. (2016a) (i.e., Gea 2016). Panel (d) presents the corresponding inverted $\boldsymbol{D_{s0}}$ profiles along with $\boldsymbol{D_{min}}$ obtained from the empirical formula by Darendeli (2001). The ranges reported below the figure indicate the minimum and maximum total misfit values obtained for each inversion parameterization.

Both the dispersion and attenuation targets are reproduced well by the predominant-mode across the investigated frequency range. Note that, for this site the predominant mode is essentially the fundamental mode. The inverted $V_S$ profiles are shown in Fig. 8c together with available invasive measurements, including downhole (DH), crosshole (CH), and seismic dilatometer (SDMT) data, as well as the median non-invasive profile reported by Garofalo et al. (2016a). Despite variations among the different layer parameterizations, all inversion results consistently reproduce the key features of the subsurface velocity structure. In particular, the profiles capture the increase in shear-wave velocity with depth and show good overall agreement with the available reference $V_S$ measurements. The corresponding damping ratio profiles are presented in Fig. 8d. A relatively consistent damping structure emerges across all parameterizations. The near-surface layers are characterized by damping ratios generally between 2% and 3%, while the deeper deposits exhibit lower damping values, typically ranging between 1% and 2%. The gradual decrease in damping with depth is broadly consistent with the increasing stiffness observed in the inverted $V_s$ profiles. For comparison, a minimum damping ratio profile was estimated using the Darendeli (2001) empirical relationship. Owing to the lack of detailed geotechnical information required by the Darendeli formulation, representative values of OCR and PI were assumed for the Mirandola site. Based on the borehole information, the subsurface was approximated by three layers corresponding to silt/fine sand (0–4 m), clay (4–10 m), and sand (10–25 m). PI of 10, 40, and 5 were assigned to the respective layers, while an OCR=1 was assumed throughout the profile. A loading frequency of 1 Hz was assumed, which is commonly adopted for developing non-linear modulus reduction and damping curves for seismic studies. The groundwater table depth of approximately 4 m was considered in the effective stress calculations. As shown in Fig. 8d, the empirical damping profile predicts lower damping values than those obtained from the attenuation-constrained inversion, although the two approaches are in close agreement within the deeper sand layer. This trend is consistent with expectations, as the in-situ attenuation values include contributions from both intrinsic damping and scattering, while the lab-based empirical relationship of Darendeli (2001) only accounts for intrinsic damping. Nevertheless, the general agreement between the empirical estimates and the inversion-derived damping values provides confidence that the inverted damping profiles are physically reasonable. Overall, the inversion results demonstrate that the proposed joint inversion framework is capable of simultaneously recovering stiffness and plausible damping profiles from field measurements.

### 7.2. Site B: Bird's Landing

The Bird's Landing site is located along the Sacramento River on Sherman Island, California, USA. The site is predominantly underlain by organic clay overlying a thick peat deposit (Bhaumik and Cox 2026b; Panthi and Cox 2026). Data acquisition was performed using 24 vertical-component geophones spaced at 2-m intervals. Active-source wavefields were generated using a sledgehammer with shot locations at 5, 10, 15, and 20 m on both sides of the receiver array. Five shots at each location were stacked to enhance the signal-to-noise ratio. An example shot gather, together with the corresponding dispersion and attenuation images, are presented in Fig. 9. The dispersion image shows many higher modes at high frequencies and consistently trends to lower phase velocities with decreasing frequency. This dispersion trend is consistent with an embedded low velocity layer, which in this case consists of the thin and relatively stiff clay layer overlying extremely soft peat deposits that extend to significant depths. The dispersion image has a clearly defined fundamental mode that extends between approximately 1.25 – 3.5 Hz (Fig. 9b). Although such coherent low frequencies from an active-source are extremely rare, the abnormally low frequency content corresponds to wavelengths that are only 18-24 m, given the exceptionally low phase velocities of 22 – 30 m/s. The attenuation image (Fig. 9c) also shows a complex trend with discontinuities at the same frequencies where modal transitions occur in the phase velocity data.

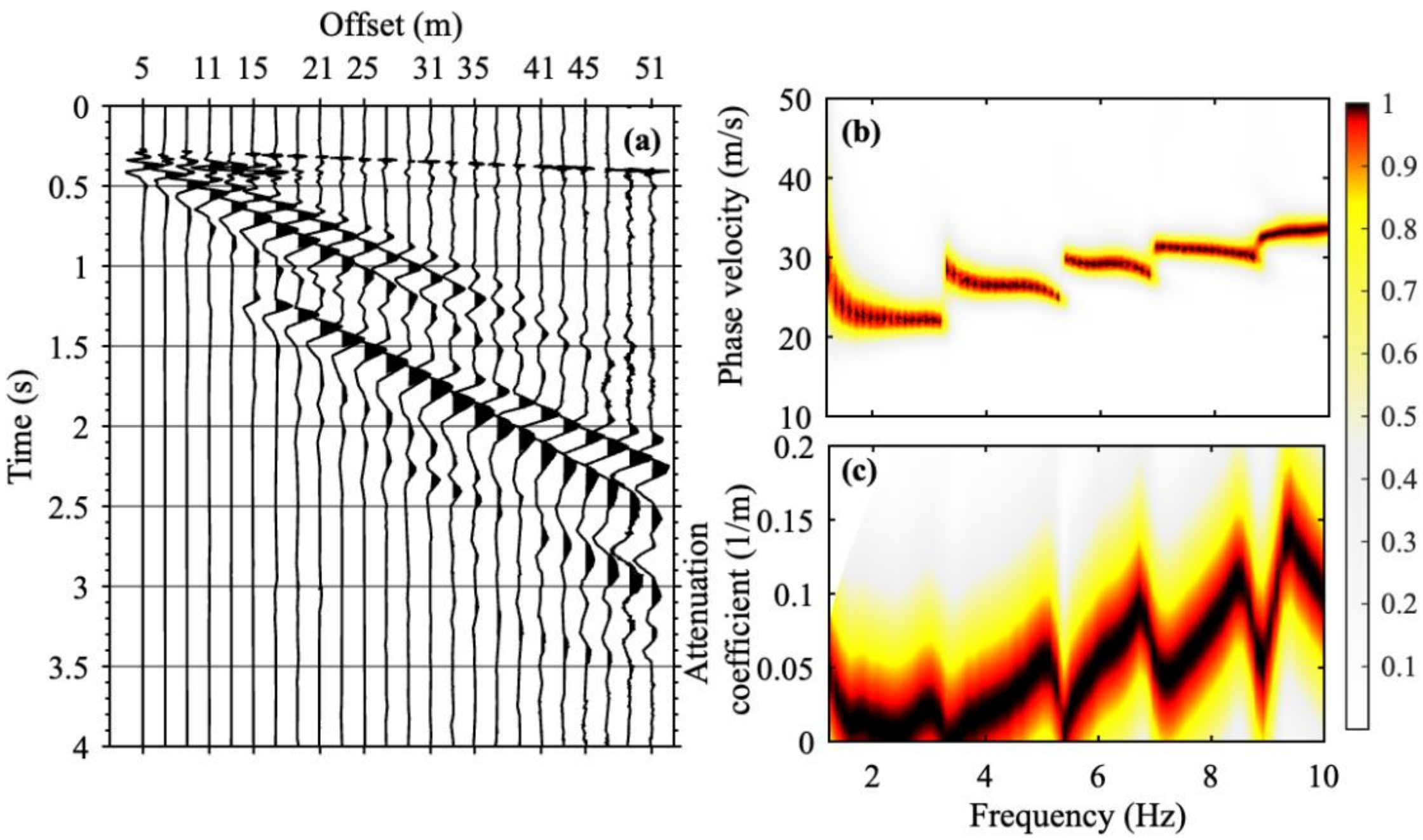

**Fig. 9.** Example 24-channel active shot waveforms for dispersion and attenuation processing at Site B: (a) recorded waveforms from a representative sledgehammer shot at a 5-m offset, (b) corresponding Rayleigh-wave dispersion image, and (c) attenuation image.

By combining the dispersion and attenuation measurements extracted from all shot locations on both sides of the array, and by removing data points falling outside the theoretical array resolution limits, a unified set of inversion targets was constructed. The resulting dispersion and attenuation target data, spanning the frequency range of approximately 1.25–10 Hz, are presented in Figs. 10a and 10b, respectively. Consistent with the previous examples, the inversion was performed using LN values of 4, 5, and 6. Both the phase velocity and attenuation targets were reproduced well by the predominant-mode inversions, indicating that the proposed framework successfully captures the dominant wavefield characteristics without requiring explicit modal identification. The five best inverted $V_S$ profiles from each of the five trials from individual layering parameterization are plotted in Fig. 10c together with the available borehole log. It should be noted that the borehole was not located directly at the test site, but approximately 100 m away; therefore, some differences between the geophysical and geotechnical observations are expected due to spatial variability. All inverted $V_S$ profiles consistently show a relatively stiff top layer followed by a thick, low-velocity layer. The low-velocity layer continues to a depth of approximately 8–10 m and is interpreted as the thick peat deposit in the borehole log. Below this depth, the shear-wave velocity gradually increases to values between 50 and 100 m/s, suggesting the presence of a relatively stiffer organic clay or clayey stratum consistent with the boring log. The corresponding damping profiles obtained from the inversion are shown in Fig. 10d. A consistent damping trend emerges across all parameterizations. The upper layer exhibits damping ratios generally ranging from approximately 2.5 to 3.5%, while the underlying peat deposit is characterized by damping values of approximately 2.5%, with the underlying clay layer predominantly between 0.5 and 2%. For comparison, the $D_{\min}$ profile estimated using the Darendeli (2001) empirical relationship is also shown in Fig. 10d. The empirical $D_{\min}$ profile was constructed using layer boundaries located at depths of approximately 1.5, 8.0, and 12.8 m based on the available borehole log. The OCR=1 was assumed for all layers, while PI values of 20, 140, and 35 were assigned to the layers based on the average measurements obtained from samples within each stratum. Representative total unit weights of 18, 12, and 18 kN/m³ were assumed for the three layers, with ground water table at around 0.8 m depth, and mean effective confining pressure was calculated assuming a coefficient of earth pressure at rest $K_0 = 0.5$. $D_{\min}$ was evaluated at the mid-depth of each layer (0.76, 4.7 and 10.4 m). It should be noted that the Darendeli (2001) relationship was not developed for highly organic soils such as peat, and the assigned PI value for the peat layer lies well outside the range of the database used for its calibration. Consequently, the corresponding estimate represents an extrapolation and should be interpreted with caution. The empirical $D_{\min}$ estimates are lower than the inverted damping values in the uppermost layer, whereas they are higher in the underlying peat layer. The discrepancy in the peat layer is consistent with the limitations of the Darendeli (2001) relationship for highly organic soils, as discussed above, and illustrate the value in being able to invert for in-situ $D_{\min}$ values for atypical and/or difficult to sample soils. The attenuation-constrained inversion directly incorporates field observations and therefore provides site-specific estimates of damping. Overall, the inversion results demonstrate the robustness of the proposed predominant-mode framework for simultaneous estimation of $V_s$ and $D_{s0}$ from field data exhibiting multiple higher modes, eliminating the need for manual modal interpretation.

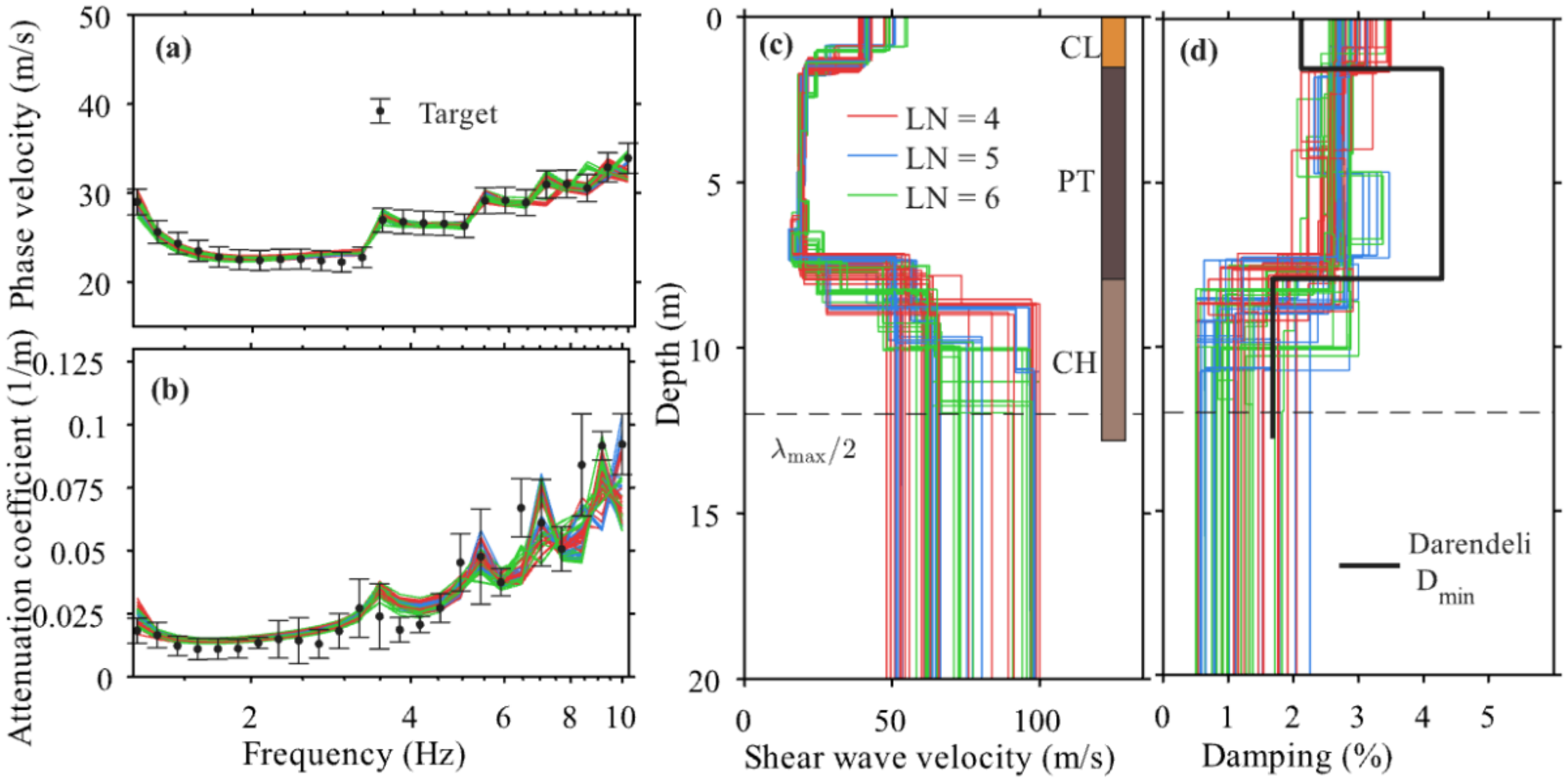

**Fig. 10.** Results of the proposed joint inversion framework for Site B. Panels (a) and (b) show the target dispersion and attenuation data together with the predominant-mode responses of the five best-fitting solutions from each inversion run (five runs for each layer parameterization: LN = 4, 5, and 6), resulting in 75 solutions in total. Panel (c) presents the inverted $\boldsymbol{V_s}$ profiles, along with the nearby borehole log. Panel (d) presents the corresponding inverted $\boldsymbol{D_{s0}}$ profiles along with $\boldsymbol{D_{\min}}$ obtained from the empirical formula by Darendeli (2001). The ranges reported below the figure indicate the minimum and maximum total misfit values obtained for each inversion parameterization.

# 8. Conclusion

This study presented a joint inversion framework for the simultaneous estimation of shear-wave velocity and damping profiles from surface-wave phase velocity and attenuation measurements. The proposed methodology addresses two fundamental limitations of conventional surface-wave inversion approaches. First, unlike traditional methods that primarily focus on stiffness characterization, the proposed framework employs a viscoelastic forward model to simultaneously estimate both stiffness and energy dissipation characteristics of the subsurface. Second, the predominant mode formulation automatically identifies the physically consistent theoretical mode, eliminating the need for subjective analyst judgment and manual mode indexing, particularly for attenuation measurements where modal identification is often challenging.

The proposed framework was implemented by extending the Thin Layer Method (TLM)-based elastic forward model to a viscoelastic formulation capable of simultaneously calculating predominant mode phase velocity and attenuation coefficients. The predominant mode attenuation formulation was subsequently validated using published attenuation curves derived from wavefield simulations for two representative synthetic models: a monotonic increasing velocity profile and a low-velocity-layer profile. In both cases, the predominant attenuation showed good agreement with the attenuation extracted from the simulated wavefields, validating and providing confidence in its application in the inversion framework. The performance of the proposed joint inversion framework was subsequently evaluated using the same synthetic models. Multiple parameterization schemes based on different numbers of layers were considered to investigate the influence of model parameterization. The resulting velocity profiles showed good agreement across different parameterizations, and the inversions successfully reconstructed both the true velocity and damping profiles. Notably, for the second model, which exhibited multiple modes, the predominant mode formulation successfully identified the correct modes without requiring explicit mode indexing. Consistent damping trends were also observed among the different parameterizations.

The applicability of the proposed framework was further demonstrated using two field datasets: Site A from Mirandola, Italy, and Site B from Bird's Landing, California. Dispersion and attenuation measurements were extracted from the experimental waveforms using the FDBFA (Frequency Domain Beam Forming Attenuation) approach and were subsequently utilized as inversion targets. Site A was characterized by predominantly fundamental-mode dispersion trends within the analyzed frequency range, whereas Site B exhibited significant higher modes due to a low-velocity peat layer. For both sites, the predominant mode responses of the inverted profiles reproduced the experimental dispersion and attenuation trends, including the apparent modal transition at Site B. The inverted shear-wave velocity profiles also showed good agreement with independent site characterization results. The estimated damping profiles were generally consistent with empirical predictions from Darendeli (2001), except within the Site B peat layer where the empirical relationship lies outside its intended range of application, while also revealing site-specific variations that cannot be captured by generalized empirical relationships alone, highlighting the value of attenuation measurements for direct characterization of subsurface damping properties. Overall, the proposed joint inversion framework provides a practical and effective approach for simultaneous characterization of subsurface stiffness and damping properties using non-invasive surface-wave measurements. Such information is particularly important for seismic site response analyses, dynamic soil-structure interaction studies, and other geotechnical applications where both stiffness and damping govern system performance. By integrating phase velocity and attenuation information within a unified predominant mode inversion strategy, the methodology advances beyond conventional velocity-only approaches or joint inversion approaches that require manual mode indexing and contributes toward more realistic characterization of geomaterials for geotechnical and earthquake engineering applications.

# 9. References

Abbas, A., M. Aimar, B. R. Cox, and S. Foti. 2025. “A Frequency-Domain Beamforming Procedure for Extracting Rayleigh Wave Attenuation Coefficients and Small-Strain Damping Ratio from 2D Ambient Noise Array Measurements.” *Earthquake Spectra*, 41 (2): 1333–1363. https://doi.org/10.1177/87552930241304914.

Aimar, M., S. Foti, and B. R. Cox. 2024a. “Novel techniques for in situ estimation of shear-wave velocity and damping ratio through MASW testing – I: a beamforming procedure for extracting Rayleigh-wave phase velocity and phase attenuation.” *Geophysical Journal International*, 237 (1): 506–524. Oxford Academic. https://doi.org/10.1093/GJI/GGAE051.

Aimar, M., S. Foti, and B. R. Cox. 2024b. “Novel techniques for in situ estimation of shear-wave velocity and damping ratio through MASW testing part II: a Monte Carlo algorithm for the joint inversion of phase velocity and phase attenuation.” *Geophysical Journal International*, 237 (1): 525–539. Oxford Academic. https://doi.org/10.1093/GJI/GGAE050.

Barbosa, J. M. de O., and E. Kausel. 2012. “The thin-layer method in a cross-anisotropic 3D space.” *International Journal for Numerical Methods in Engineering*, 89 (5): 537–560. https://doi.org/10.1002/nme.3246.

Bergamo, P., S. Maranò, and D. Fäh. 2023. “Joint estimation of S-wave velocity and damping ratio of the near-surface from active Rayleigh wave surveys processed with a wavefield decomposition approach.” *Geophysical Journal International*, 233 (3): 1560–1579. Oxford Academic. https://doi.org/10.1093/gji/ggad010.

Bhaumik, M., and B. R. Cox. 2026a. “Predominant-mode inversion of surface waves: Inherently addressing inconspicuous low frequency mode jumps.” *Engineering Geology*, 369: 108834. https://doi.org/10.1016/j.enggeo.2026.108834.

Bhaumik, M., and B. R. Cox. 2026b. “Radial-Component Predominant-Mode Inversion of Rayleigh Waves: Application to DAS-Based Site Characterization.” *Journal of Geophysical Research: Solid Earth*, 131 (10): e2026JB034811. https://doi.org/10.1029/2026JB034811.

Bhaumik, M., and T. Naskar. 2024a. “Higher-order thin layer method as an efficient forward model for calculating dispersion curves of surface and Lamb waves in layered media.” *Computers and Geosciences*, 186: 105570–105570. https://doi.org/10.1016/j.cageo.2024.105570.

Bhaumik, M., and T. Naskar. 2024b. “Computation of Surface Wave’s Dominating Mode For Stratified Media.” *Indian Geotechnical Journal*, 1–16. Springer. https://doi.org/10.1007/s40098-024-01045-x.

Boaga, J., G. Cassiani, C. L. Strobbia, and G. Vignoli. 2013. "Mode misidentification in Rayleigh waves: Ellipticity as a cause and a cure." *Geophysics*, 78 (4): EN17–EN28. https://doi.org/10.1190/geo2012-0194.1.

Cox, B. R., P. Panthi, K. Cannon, and M. Bhaumik. 2026. "Distributed Acoustic Sensing (DAS) and Nodal Station Measurements of the Birds Landing Transmission Tower Implosion." DesignSafe-CI.

Cox, B. R., and D. P. Teague. 2016. "Layering ratios: A systematic approach to the inversion of surface wave data in the absence of a priori information." *Geophysical Journal International*, 207 (1): 422–438. Oxford Academic. https://doi.org/10.1093/gji/ggw282.

Darendeli, M. B. 2001. "Development of a new family of normalized modulus reduction and material damping curves." The University of Texas at Austin.

Dawadi, N., K. Mohammadi, M. M. Hallal, and B. R. Cox. 2026. "Insights on Numerical Damping Formulations Gained From Calibrating 2D Ground Response Analyses at Downhole Array Sites." *Earthquake Spectra*, 42 (3): e70087. https://doi.org/10.1002/esp4.70087.

Foti, S., F. Hollender, F. Garofalo, D. Albarello, M. Asten, P.-Y. Bard, C. Comina, C. Cornou, B. Cox, G. Di Giulio, T. Forbriger, K. Hayashi, E. Lunedei, A. Martin, D. Mercerat, M. Ohrnberger, V. Poggi, F. Renalier, D. Sicilia, and V. Socco. 2018. "Guidelines for the good practice of surface wave analysis: a product of the InterPACIFIC project." *Bulletin of Earthquake Engineering*, 16 (6): 2367–2420. https://doi.org/10.1007/s10518-017-0206-7.

Garofalo, F., S. Foti, F. Hollender, P. Y. Bard, C. Cornou, B. R. Cox, A. Dechamp, M. Ohrnberger, V. Perron, D. Sicilia, D. Teague, and C. Vergniault. 2016a. "InterPACIFIC project: Comparison of invasive and non-invasive methods for seismic site characterization. Part II: Inter-comparison between surface-wave and borehole methods." *Soil Dynamics and Earthquake Engineering*, 82: 241–254. Elsevier. https://doi.org/10.1016/j.soildyn.2015.12.009.

Garofalo, F., S. Foti, F. Hollender, P. Y. Bard, C. Cornou, B. R. Cox, M. Ohrnberger, D. Sicilia, M. Asten, G. Di Giulio, T. Forbriger, B. Guillier, K. Hayashi, A. Martin, S. Matsushima, D. Mercerat, V. Poggi, and H. Yamanaka. 2016b. "InterPACIFIC project: Comparison of invasive and non-invasive methods for seismic site characterization. Part I: Intra-comparison of surface wave methods." *Soil Dynamics and Earthquake Engineering*, 82: 222–240. Elsevier. https://doi.org/10.1016/j.soildyn.2015.12.010.

Kausel, E., P. Malischewsky, and J. Barbosa. 2015. "Osculations of spectral lines in a layered medium." *Wave Motion*, 56: 22–42. Elsevier B.V. https://doi.org/10.1016/j.wavemoti.2015.01.004.

Kausel, E., and R. Peek. 1982. "Dynamic loads in the interior of a layered stratum: An explicit solution." *Bulletin of the Seismological Society of America*, 72 (5): 1459–1481. GeoScienceWorld. https://doi.org/10.1785/BSSA0720051459.

Kausel, E., and J. M. Roësset. 1981. "Stiffness matrices for layered soils." *Bulletin of the Seismological Society of America*, 71 (6): 1743–1761. https://doi.org/10.1785/BSSA0710061743.

Kokusho, T. 2017. *Innovative Earthquake Soil Dynamics*. London: CRC Press.

Kramer, S. L., and J. P. Stewart. 2024. *Geotechnical Earthquake Engineering*. CRC Press.

Lai, C. G. 1998. "Simultaneous inversion of Rayleigh phase velocity and attenuation for near-surface site characterization." Georgia Institute of Technology.

Lai, C. G., and G. J. Rix. 2002. "Solution of the Rayleigh Eigenproblem in Viscoelastic Media." *Bulletin of the Seismological Society of America*, 92 (6): 2297–2309. https://doi.org/10.1785/0120010165.

Lai, C. G., G. J. Rix, S. Foti, and V. Roma. 2002. "Simultaneous measurement and inversion of surface wave dispersion and attenuation curves." *Soil Dynamics and Earthquake Engineering*, 22 (9–12): 923–930. Elsevier BV. https://doi.org/10.1016/S0267-7261(02)00116-1.

Leeuw, J. de, K. Hornik, and P. Mair. 2010. "Isotone Optimization in R: Pool-Adjacent-Violators Algorithm (PAVA) and Active Set Methods." *Journal of Statistical Software*, 32: 1–24. https://doi.org/10.18637/jss.v032.i05.

Menq, F. 2003. "Dynamic properties of sandy and gravelly soils." The University of Texas at Austin. http://hdl.handle.net/2152/779.

Meza-Fajardo, K. C., and C. G. Lai. 2007. "Explicit causal relations between material damping ratio and phase velocity from exact solutions of the dispersion equations of linear viscoelasticity."

*Geophysical Journal International*, 171 (3): 1247–1257. https://doi.org/10.1111/j.1365-246X.2007.03590.x.

Nazarian, S., K. H. Stokoe, and W. R. Hudson. 1983. "Use of spectral analysis of surface waves method for determination of moduli and thicknesses of pavement systems." *Transportation Research Record*, 38–45.

Panthi, P., and B. R. Cox. 2026. "Comparative Analysis of Strain-to-Velocity Conversion Methods for Active-Source DAS Data and Collocated Nodal Stations." *Sensors*, 26 (15): 4673. Multidisciplinary Digital Publishing Institute. https://doi.org/10.3390/s26154673.

Park, C. B., R. D. Miller, and J. Xia. 1999. "Multichannel analysis of surface waves." *Geophysics*, 64 (3): 800–808. Soc of Exploration Geophysicists. https://doi.org/10.1190/1.1444590.

Parolai, S., C. G. Lai, I. Dreossi, O. J. Ktenidou, and A. Yong. 2022. "A review of near-surface QS estimation methods using active and passive sources." *Journal of Seismology*, 26 (4): 823–862. Springer Science and Business Media B.V. https://doi.org/10.1007/s10950-021-10066-5.

Rix, G. J., C. G. Lai, and A. W. Spang. 2000. "In Situ Measurement of Damping Ratio Using Surface Waves." *Journal of Geotechnical and Geoenvironmental Engineering*, 126 (5): 472–480. American Society of Civil Engineers (ASCE). https://doi.org/10.1061/(ASCE)1090-0241(2000)126:5(472).

Storn, R., and K. Price. 1997. "Differential Evolution - A Simple and Efficient Heuristic for Global Optimization over Continuous Spaces." *Journal of Global Optimization*, 11 (4): 341–359. Springer Netherlands. https://doi.org/10.1023/A:1008202821328.

Thomson, W. T. 1950. "Transmission of Elastic Waves through a Stratified Solid Medium." *Journal of Applied Physics*, 21 (2): 89–93. https://doi.org/10.1063/1.1699629.

Vantassel, J. P., and B. R. Cox. 2021. "SWinvert: A workflow for performing rigorous 1-D surface wave inversions." *Geophysical Journal International*, 224 (2): 1141–1156. Oxford University Press. https://doi.org/10.1093/gji/ggaa426.

Verachtert, R., G. Lombaert, and G. Degrande. 2018a. "Multimodal determination of Rayleigh dispersion and attenuation curves using the circle fit method." *Geophysical Journal International*, 212 (3): 2143–2158. Oxford University Press. https://doi.org/10.1093/GJI/GGX523.

Verachtert, R., G. Lombaert, and G. Degrande. 2018b. "Sensitivities of the Rayleigh and Love Phase Velocities and Attenuation Coefficients." *Journal of Geotechnical and Geoenvironmental Engineering*, 144 (11). American Society of Civil Engineers (ASCE). https://doi.org/10.1061/(ASCE)GT.1943-5606.0001978.

Vucetic, M. 1994. "Cyclic Threshold Shear Strains in Soils." *Journal of Geotechnical Engineering*, 120 (12): 2208–2228. American Society of Civil Engineers. https://doi.org/10.1061/(ASCE)0733-9410(1994)120:12(2208).

Wathelet, M. 2005. "Array recordings of ambient vibrations : surface-wave inversion." *Liège University (Belgium)*.

Xia, J., R. D. Miller, and C. B. Park. 1999. "Estimation of near-surface shear-wave velocity by inversion of Rayleigh waves." *Geophysics*, 64 (3): 691–700. Soc of Exploration Geophysicists. https://doi.org/10.1190/1.1444578.

Zhang, J., and A. C. Sanderson. 2009. "JADE: Adaptive differential evolution with optional external archive." *IEEE Transactions on Evolutionary Computation*, 13 (5): 945–958. https://doi.org/10.1109/TEVC.2009.2014613.

Zywicki, D. J. 1999. "Advanced signal processing methods applied to engineering analysis of seismic surface waves." Georgia Institute of Technology.